# Measuring relative opinion from location-based social media:

# A case study of the 2016 U.S. presidential election

Zhaoya Gong[1]*, Tengteng Cai[2], Jean-Claude Thill[2], Scott Hale[3], Mark Graham[3]

1. School of Geography, Earth and Environmental Sciences, University of Birmingham, UK
2. Public Policy Program, University of North Carolina at Charlotte, USA
3. Oxford Internet Institute, University of Oxford, UK

*Corresponding author

Email: z.gong@bham.ac.uk

# Abstract

Social media has become an emerging alternative to opinion polls for public opinion collection, while it is still posing many challenges as a passive data source, such as structurelessness, quantifiability, and representativeness. Social media data with geotags provide new opportunities to unveil the geographic locations of users expressing their opinions. This paper aims to answer two questions: 1) whether quantifiable measurement of public opinion can be obtained from social media and 2) whether it can produce better or complementary measures compared to opinion polls. This research proposes a novel approach to measure the relative opinion of Twitter users towards public issues in order to accommodate more complex opinion structures and take advantage of the geography pertaining to the public issues. To ensure that this new measure is technically feasible, a modeling framework is developed including building a training dataset by adopting a state-of-the-art approach and devising a new deep learning method called Opinion-Oriented Word Embedding. With a case study of tweets on the 2016 U.S. presidential election, we demonstrate the predictive superiority of our relative opinion approach and we show how it can aid visual analytics and support opinion predictions. Although the relative opinion measure is proved to be more robust than polling, our study also suggests that the former can advantageously complement the latter in opinion prediction.

# Introduction

Measuring and monitoring public opinion trends from social media has emerged as a potential alternative to opinion polls due to its voluntary nature and penetration to a large number of people [1]. Almost all social media platforms (e.g., Twitter and Facebook) allow users to tag their locations on the posted messages, dubbed Location-based social media (LBSM). Bringing a geographic perspective allows the study of opinion variation across geographic entities (e.g., states) that frame public events, e.g., political elections.

However, as an organic and passive data source, social media data pose several analytical challenges such as how to identify the target information from the unstructured and unprompted data, how to quantify the highly qualitative textual messages, and how to ensure the data can be representative of the broader electorate. Hence, two fundamental concerns need to be grappled with: 1) whether quantifiable measurement of public opinion can be garnered reliably from social media and 2) whether it can produce better or complementary measures compared to opinion polls [2,3].

Also, the practice of opinion polling has intrinsic limitations. All polls measure opinion in an absolute sense, where opinions are classified as one of several predefined and mutually exclusive categories, such as candidates A, B, and C. This way is conducive to overlooking complex opinion structures that would be embedded in an opinion space comprising every category as its own dimension. With each dimension featuring a gradient of preference level for one category (e.g., a range from anti-A to pro-A), some very complex opinion position could be triangulated from these dimensions, such as not very anti-A, somewhat pro-B, but no preference to C.

This research aims to advance the measurement of public opinion captured from Twitter posts, by addressing several of the points raised above. First, a relative opinion measure is proposed, based on a concept of relational space framed by the modalities of functional relationships between entities (e.g., individual persons or geographic areas). It enables the construction of a multi-dimensional and continuous representation of opinion space to 1) account for complex opinion structures arising from discrete extremes/categories (e.g. swing states in U.S. presidential election) and 2) encompass sufficient dimensions that individually characterize the opinion space from a certain aspect. Second, relative opinion positions of Twitter users are learned from textual tweets and represented as points in the multi-dimensional opinion space. A novel deep learning model known as opinion-oriented word embedding is devised to learn vector representations of words from a corpus of textual posts whose opinion indication is clearly captured by a set of selected hash tags. Third, the power of the relative opinion measure is twofold: 1) creating a spatial visualization of the opinion space where users' opinion positions can be aggregated to any level of geography based on users' location information; 2) supporting opinion predictions at an aggregate geographic level consistent with the target public event (e.g., the state level for presidential elections) via a linear neighborhood propagation method.

The rest of this paper is structured as follows. The next section reviews the measurement of public opinion with regard to methodological approaches and data sources. The conceptualization and construction of relative opinion measure is then advanced with a general modeling framework. It is followed by a detailed description and explanation of data collection and methods supporting the implementation of this framework. With an application to tweets during the 2016 U.S. presidential election, the implementation demonstrates the spatial visualization of the relative opinion space for visual analytics; the following section extends this work to opinion predictions in comparison to opinion polls. Finally, conclusions are drawn on the scientific merit of the relative opinion measure and future work is discussed.

## Literature review

# Social media data as an alternative source for opinion measurement

Public opinion consists of people's beliefs, attitudes and desires on public issues or problems. For governments, political leaders, and policy makers, discerning public opinion is crucial to inform administration, election campaign, and policy making [4]. Data are traditionally collected by survey or opinion poll. These techniques involve a structured questionnaire, a defined population from which individuals are sampled, and a method of aggregation of individual responses to infer a quantity of interest (as a measure of opinion) for the population, as core components [2,5].

With the proliferation of web and mobile technologies, social media platforms, such as Twitter and Facebook, have permeated large segments of population worldwide; they let people express their thoughts, feelings, attitudes and opinions, which can be shared and accessed publicly [4]. This presents not only a ubiquitous means to convey individual opinion but also an unprecedented alternative source of public opinion data. In contrast to surveys, this new form of public opinion data is characterized by unstructured and unprompted opinion expression. In essence, it belongs to a type of organic or passive data that users voluntarily post on social media. That said, social media data can prevent the prompting or framing effects that may exist in surveys when respondents and their responses are oriented and affected by how the questionnaire designers select and frame the topics/issues [2,5–7].

Moreover, social media data have unparalleled advantages in temporal and geographic coverages at very fine granularity. Users across countries and from different geographic regions may post on a daily or even hourly basis. Indeed, it is likely to capture people's instantaneous and spontaneous responses to public events and issues and their changes over time, which is impossible with survey data because of the cost and practicality [2,8]. Almost all social media platforms allow users to tag their locations on the posted messages. At the finest level, the exact location such as a pair of geographic coordinates can be reported, although larger geographic regions are more common such as towns, cities or states. Geographic variation of public opinions plays a critical role in many situations such as electoral-area-based elections (e.g., congressional district and state). Due to the cost of opinion polls, social media data have been proposed to interpolate state-level polls for U.S. presidential elections [9,10]. The timeliness and geographic reach of social media data reinforce their appeal as opinion polls are facing growing hurdles in reaching and persuading reluctant respondents [2].

## Challenges of LBSM data for opinion measurement

Measuring public opinion with LBSM data is still confronting several main challenges. First, given the unstructured and unprompted nature of social media data, how to determine the topics and relevant posts from a huge pool of social media data is a great challenge. It has been argued that simple ad hoc search criteria, such as the mention of candidate names on the ballot, may cause systematic selection bias. Specifically, this approach may miss those relevant messages that fail to mention candidates by name or add noise to the data when candidate names happen to be confused with other names [2,11]. Therefore, the selection criteria need to be carefully thought through and the potential selection bias needs to be assessed with the interpretation of results. The difficulty of identifying topics also lies in that topics are changing, they are related to one another or split into sub-topics as discourse is carried on over time [2]. Efforts have been made to discover related topics or sub-topics either explicitly by using a combination of topic models and sentiment analysis [12] or implicitly by constructing a network of co-occurrent hashtags referring to related topics [3].

Second, quantification of opinion from qualitative and unstructured textual data posted on social media is not only a technical challenge but also a theoretical one. Simple metrics based on counting tweets or mentions related to certain topics/issues or candidates/parties, though widely applied, have been criticized for their low performance compared to real outcomes or opinion polls [13–17]. As an enhancement to simple counting methods, lexicon-based sentiment analysis has been used in numerous studies to extract positive or negative sentiments from textual messages on certain topics [2,14,16,18]. However, both types of methods are in fact measuring attentions or sentiments rather than opinions (e.g., attitude regarding an argument or preference for a candidate or a party) [10,16,19,20]. Moreover, the lexicon-based approach often exhibits unstable performance on the unstructured, informal, sometimes ironic or sarcastic, language of social media messages due to an ad-hoc dictionary of words with sentiment polarity it relies on [21].

Recent research has built more accurate measurements of opinion by taking a supervised learning approach with either a manually created or automatically generated in-domain (instead of ad-hoc dictionary) training set that identifies exact opinion information, such as political support/opposing or agreement/disagreement [3,16,19,22]. Taking a 'bag-of-words' approach in natural language processing (NLP) [23], these studies all assume that every word of a message constitutes a piece of the opinion expressed by the message as a whole, whether this word is directly related to the topic or considered neutral. These stimulating attempts to transform qualitative textual data into quantitative measures of opinion are however very elementary due to the nature of bag-of-words representation (a vector of word appearance) taking no consideration of word order and hence no semantic information captured at the word level.

Recent advance in deep neural network learning of word representation (or word embedding) as dense, low-dimensional and real-valued vectors, has suggested superior performance compared to bag-of-words-based methods. The neural-network-inspired word embedding has been proved effective in a variety of NLP tasks including sentiment analysis [24–28]. Formally, word embedding maps words and phrases to mathematic vectors, creating quantitative representations in a multidimensional space that preserve syntactic, semantic and contextual relationships of words [29]. For instance, the well-known word2vec model [25] can achieve tasks like king – queen = man – woman or infer pairs of present tense-past tense. However, there is limited research incorporating semantic-preserving word embedding for opinion measurement (some studies using it for topic detection only, such as [30,31]).

Third, there is no formal process for defining a population frame and drawing samples from social media data as in a survey. The representativeness of social media data is questionable, although the decentralized nature of this data source and the diversity of its users may compensate for the potential bias, owing to the large size of social media data [2]. A number of studies has shown that social media users are not representative of the national populations in many aspects, such as their geographical distribution, age, gender, race, educational level, political ideology, and interests in topics [32–37]. For example, geotagged Twitter users in the U.S. are more likely to be younger, have higher income, live in urbanized areas, and be located in the east or west coastal areas [34]. It was also found that the majority of Twitter users are female, but they are not politically active [35]. Besides, measuring public opinion from social media depends on users who publicly express their opinions. However, these active participants who voluntarily offer opinions may have systematically different opinions on a topic from those who are explicitly asked or choose not to offer their opinions (e.g., shy

Trump voter issue); hence the former users' opinions are over-represented in social media. The underlying uncertainties are really hard to control, without even mentioning the problems of bots, spammers and fake user accounts [38]. Further research has been called for to assess the extent of uncertainty involved and how plausible social media data can be used as a trustworthy source of opinion measurement.

The representativeness issue also plays a critical role in aggregating information from individual social media messages to a measure of public opinion. For example, a small group of users act as opinion leaders and dominate the discussion on social media in terms of the volume of tweets or retweets [39]. User level effect must be controlled for to avoid the over-representation of high-level participation. Furthermore, the geotagging of social media posts may enable the aggregation to a certain geography when electoral district-based opinion measurement is necessary. However, the resolution of the tagged geographic information varies significantly cross messages (e.g., tweets), with only a small proportion of them having exact locations [40]. Geotags are volunteered by users and hence selecting only geotagged tweets may introduce a selection bias, which again causes the representativeness problem [41].

## Contributions of this work

This research responds to the second challenge and proposes a novel approach for opinion measurement, which is both theoretically grounded and technically tailored to location-based social media data. Supervised learning ensures the measurement of opinion rather than attention or sentiment. Thanks to semantic-preserving word embedding, it also ensures the capture of opinion information at the finest grain (i.e., word level). Thus, our measure is flexible enough for aggregation at a range of levels, such as message, user, and various levels of geographic granularity. Furthermore, this study partially addresses the third challenge by producing spatial representations of opinion measures at an aggregated level, which permits a straightforward assessment of representativeness of the social media data. In addition, to account for the topic selection issue posed by the first challenge, we employ the topic discovery and opinion identification methods proposed by [3] to build a training set for supervised learning.

However, this work differentiates itself from [3] in the following key aspects. First, we propose an innovative approach to measure relative opinion. Second, leveraging location-based social media data, location data are combined with textual data for opinion measurement in order to obtain a geographic representation of public opinion that is relevant to political elections. Third, opinion predictions are conducted at an aggregated geographic level (e.g., state) to better mitigate errors. As it is common that one may express conflicting opinions in different posts at different time and as the user retains full control of the content of a message after its initial post through changes, edits or even removal, individual opinion is never devoid of uncertainty. Evidence has shown that opinion classification error at the individual level (e.g., tweet or user) remains high and can propagate with aggregation [22,42].

## Conceptualization and construction of relative opinion

As an organic source of data for public opinion extraction, social media is characterized by unstructured text, which contrasts with the designed 'question' and 'response' structure of traditional survey data (opinion polls). Opinions may be embedded or even hidden in this unstructured and free-form writing, which is naturally fuzzy, complex, and of high dimensionality [2]. In and of themselves, dimensions in social media text are implicit, hidden and most of the time undeterminable. Hence, the opinions extracted from such free-form discourse inherit these features, which must be properly handled when taking the measurement.

While public opinions revealed from survey data can also be multi-dimensional, the dimensions are encoded explicitly in specific questions. In this sense, the opinion measurement taken by survey data is static, deterministic, and certain and may be called *absolute opinion*. This conceptualization treats the opinion measurement as a classification problem with predefined opinion categories, e.g. the support of candidates or parties [3,12] and permeates existing practices in statistical analysis of mentions and in sentiment analysis for social media data. Rooted in the absolute opinion paradigm, this conceptualization is ill-suited to capture the complexity of opinion structures and the continuity between opinion categories that may exist in the high-dimensionality free-form discourse on social media.

In response to the deficiencies of the absolute opinion approach for the measurement of opinions from social media posts, we hereafter propose a relative opinion conceptualization. This approach is inspired from the relativist view on space in physics and geography, which posits that space is not merely a neutral container for natural and socioeconomic processes, but that it is in turn defined and (re-)constructed by the relations among things and events that take place and operate within it [43,44]. Specifically, relative opinion is to measure how dissimilar one's opinion is from that of others. Once the semantic relations for every pair of individual agents' (person, community or state) opinions are captured, they can serve to frame the construction of a relative opinion space. This space entails a multidimensional and continuous representation of opinions that can account for complex opinion structures (e.g. swing states in U.S. presidential election) and sufficient dimensionality of the opinion space.

To construct the relative opinion measure from Twitter data, we further propose a modeling framework (Fig 1). It is our fundamental contention that opinion information can be incorporated into the learning process of generic word embedding to preserve the opinion orientation of words and sentences towards certain topics, such as support for candidates or parties. This framework comprises three components: 1) data collection and training data for opinion identification, 2) generation of opinion-oriented word embedding, and 3) aggregation of word-level opinion embedding to individual and higher level of geographic units (the state in the case study of this paper). The output is an aggregate relative opinion measure at the state level and can be useful for visual analytics and predictive analysis in the case study of the 2016 U.S. presidential election.

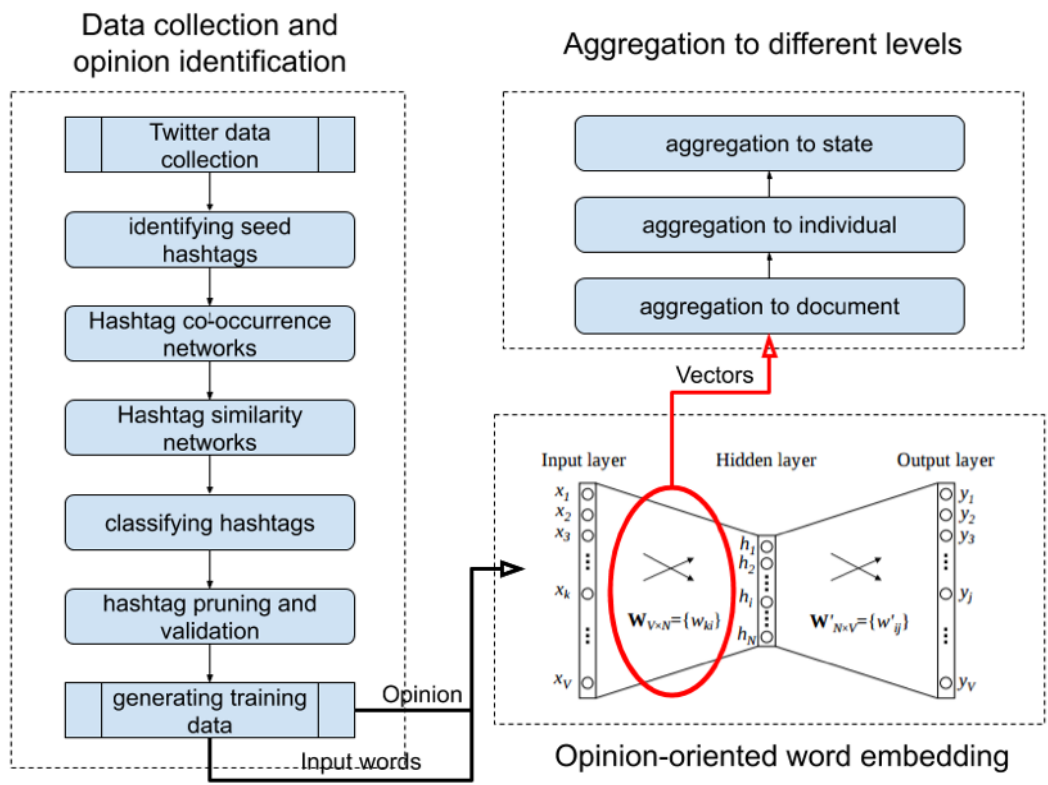

**Fig 1. Modeling framework for relative opinion**

# Methods and data

## Data collection and training data for opinion identification

We continuously collected tweets using the Twitter Streaming API (about 1% of Twitter posts) from September 1st, 2016 to November 8th, 2016, date of the 2016 U.S. presidential elections. A total of 2.2 million tweets were collected in the English language from the United States with certain location tags, mentioning the two top presidential candidates from the Republican Party (Donald J. Trump) and the Democratic Party (Hillary Clinton) by using the same filters as in [3] with the following keywords: trump OR realdonaldtrump OR donaldtrump and hillary OR clinton OR hillaryclinton. As in [3], we use the name of the Twitter client extracted from each raw tweet to filter out those automated tweets from bots. 90% of the collected tweets are retained as originating from official clients. Following the procedure in [3], we utilize hashtags from tweets as the main source of opinion information to build a training set of labelled tweets that indicate clear opinion preferences. This five-step procedure (Fig 1) is detailed in section A of the S1 Appendix. The output of the procedure is a set of labeled tweets in terms of six opinion categories: Pro-Clinton, Anti-Trump, Support-Clinton (with Pro-Clinton and Anti-Trump two most common labels), Pro-Trump, Anti-Clinton, and Support-Trump (with Pro-Trump and Anti-Clinton two most common labels). This training set contains 238,142 tweets.

## Opinion-oriented word embedding

We develop our opinion-oriented word embedding (OOWE) as an extension of the sentiment-specific word embedding (SSWE) method [26]. SSWE incorporates sentiment information (e.g., positive/negative emoticons) from tweets into generic semantics-preserving word embedding. OOWE distinguishes itself from SSWE in two respects (Fig 2): 1) it is supervised learning with opinion preference rather than sentiment information; 2) it can accommodate any number of opinion categories rather than two (positive-negative) for sentiments. Specifically, we modify the upper linear layer to include two separate components, opinion and semantic, which capture the opinion preference and semantic context of words, respectively. Given $C$ opinion categories, the output layer outputs a $C + 1$ dimensional vector in which one scalar $f_s$ stands for the language model score and $f_o^j$ ($j = 1, \ldots, C$) scalars stand for opinion scores for all categories. The loss function is specified as a linear combination of two hinge losses:

$$\begin{cases} Loss(t, t^r) = (1-\alpha) loss_s(t, t^r) + \alpha\, loss_o(t) \\ loss_s(t, t^r) = max(0, 1 + f_s(t^r) - f_s(t)) \\ loss_o(t) = \frac{1}{C-1} \sum_{j \neq P}^{C} max(0, 1 + f_o^j(t) - f_o^P(t)) \end{cases} \quad (1)$$

where $t$ and $t^r$ are original and corrupted ngram inputs, respectively, $\alpha$ is a weighting parameter, $f_o^P$ is the opinion score for the Positive opinion category while $f_o^j$ ($j \neq P$) is the opinion scores for other Negative opinion categories.

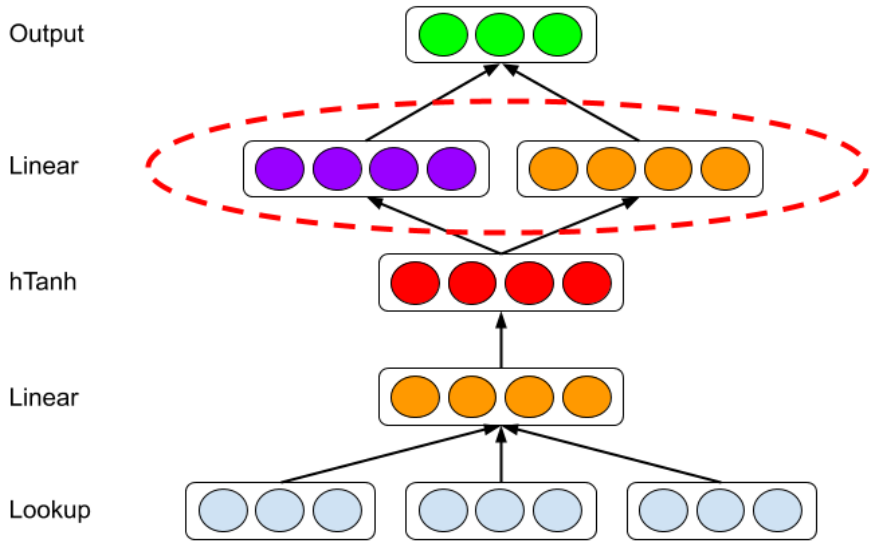

**Fig 2. Neural network structure for opinion-oriented word embedding algorithm**

We tokenized each tweet with TwitterNLP [45], remove the @user and URLs of each tweet, and hashtags in the set of labeled hashtags. We train OOWE by taking the derivative of the loss through back-propagation with respect to the whole set of parameters [29], and use AdaGrad [46] to update the parameters. We empirically set the window size as 3, the embedding length as 50, the length of hidden layer as 20, and the learning rate of AdaGrad as 0.1. After training, the output word embeddings are numeric vectors whose relative positions and distances toward opinion categories (e.g., Pro-Clinton and Anti-Trump) represent the relative opinions.

## Aggregation of relative opinion measure

Since we measure relative opinion at the word level, it is possible to aggregate it to any higher level, such as the tweet, the user, and the state. A straightforward way to aggregate the embedding representation of words to a document is to take their centroids (averages), which has been a common approach in creating document-level embedding [47]. Relative opinion measures at the user and state levels can be obtained similarly. As suggested in a recent review of ways to measure public opinion with social media data by [2], different levels of participation of users in social media, as reflected by the varying number of tweets posted by different users, should be controlled for at the user level. The evaluations of relative opinion measures at the word and user levels show that these measures can correctly capture the opinion orientation for specific hashtags and users, which is detailed in section B of the S1 Appendix. Location tags in social media allow aggregation of opinion measures by administrative and geographic areas, which may be very useful in electoral studies. The limitation of the representativeness of geotagged tweets is well recognized. We assume here that users would be willing to reveal coarser (e.g., state or country) rather than finer (e.g., coordinates or city) location information. We observe that state level location can be inferred for around 90% of the tweets we collected from one of three pieces of information: Tweet location field, mentioned location in tweet text, and user profile location field.

## Visual analytics of state-level relative opinion measures

This section demonstrates the usefulness of the relative opinion measure in visual analytics by creating a spatial visualization of the relative opinion space where state-level opinions are represented as points. Figs 3-4 are 2-dimensional representations of the original 50-dimension state-level opinion embedding. They are generated by regular Multi-Dimensional Scaling (MDS, [48]), a linear dimensionality reduction technique to construct a low-dimensionality presentation and preserve the global geometry of high-dimensionality data points. The two axes depicted in these figures are arbitrarily determined in MDS. In other words, what matters is the relative distance, representing the dissimilarity of opinions, between states as data points embedded in the constructed relative opinion space. Colors in the figures represent the 2016 U.S. presidential popular vote results by state for the top two candidates: Trump (red) and Clinton (blue). In Fig 3, states that voted for different candidates are quite clearly divided in the estimated opinion space, which demonstrates that the relative opinion measure is capable of capturing a general pattern of discriminating opinions among states. The opinion extremes follow the upper-right-bottom-left axis, with Support-Clinton heading in the bottom-left direction and Support-Trump towards the upper-right direction. It is also noted that swing states (like PA, MI, NH, NV, VA, AZ, and WI) are along, or even cross, the dividing boundary of opinion, which indicates their nature of having similar levels of support for both candidates. This is better shown by the ratio of votes for Trump to votes for Clinton with the gradient of colors in Fig 4. In this figure, there is a general trend of darker colors towards the two extremes of opinions, while the lighter colors are found close to the dividing boundary of opinions.

In addition, variation of opinions across users can be visualized for state-level opinion aggregates. The opinion embedding constructed earlier for users are taken to compute standard deviations by state as a measure of opinion variation at the user level. In Fig 5, the size of circles represents the level of opinion variation. Fig 6 visualizes the representativeness of the Twitter data by the ratio of the number of twitter users captured in the data to the size of the population of each state. Thus, the tiny circle for DC in Fig 5 indicates that users in DC share a common opinion, while the large circle of DC in Fig 6 means that DC's population is relatively well represented in the data. A small variation of opinion is also found for ME in Fig 5. However, as ME's population is relatively underrepresented in the data (a small circle in Fig 6), the small variation of opinion for ME shown in Fig 5 may in fact be biased. This is partially confirmed by the popular vote outcome where the votes casted for both candidates in ME are quite competitive, having only 2% difference (47.8% for Clinton and 45.1% for Trump).

Results show a large variation of opinions for Texas users (Fig 5) with a reasonable population representation (Fig 6). Texas cities, such as Houston, have been experiencing growth in relatively liberal urban professionals as well as in an influx of Hispanic and other immigrants, which makes urban areas swing strongly democratic [49]. In fact, in 2016, Clinton received three percent more votes than Obama did in 2012 in this state. This could explain the large variation of opinions in Texas instead, which departs from the preconceived intuition that Texas has always been a "deep-red" state. In other results, we see that AL, MS, and LA on the far extreme of the red side also show large variation (Fig 5), which indicates that opinions can differ broadly across individuals even in "deep-red" states, as republicans can come in widely different flavors. However, the former case of Texas that sits close to the dividing boundary is much more critical than the later ones for the election outcome. It is noted that Kansas is placed on the dividing boundary and carries a large opinion variation with a relatively bad representativeness. At first glance, this could be interpreted as a misplacement due to the data issue, as Kansas has always supported Republican candidates in the past four presidential elections (a deep-red state). A look into county-level election results (Table 1) point to a very different story, however. We see that the most populous county, Johnson County, had quite close supports for both candidates. This is also true for the total support of the five most populous counties. This state of affairs could be an explanation for the large variation of opinions statewide, given the commonly accepted assumption that people in more urbanized areas have a larger chance to tweet [41].

**Table 1 Election results for the top 5 most populous counties in the state of Kansas**

| County | Clinton % | Trump % | Others % | Total |
|---|---|---|---|---|
| **Johnson** | 44.76 | 47.40 | 7.84 | 290,090 |
| **Sedgwick** | 36.88 | 55.28 | 7.84 | 188,783 |
| **Shawnee** | 44.99 | 47.65 | 7.35 | 75,406 |
| **Douglas** | 62.28 | 29.32 | 8.39 | 50,087 |
| **Wyandotte** | 61.80 | 32.40 | 5.80 | 48,781 |
| **Total** | 45.13 | 47.20 | 7.67 | 653,147 |

States near the opinion boundary are interesting to consider in more detail because opinion may be harder to apprehend there than in other states. Let us now consider the case of Connecticut and Delaware, both of which have been reliably democratic states in the past four presidential elections. Records show that both states have shifted more than 5 percentage points toward the political center since 2008, reflecting President Trump's strength with white, working-class voters [50]. This shift is well captured by Twitter data as well as we find these states positioned along the dividing line, while showing a modest level of opinion variation and representativeness. In Figs 3-4, an obvious misplacement is WI which, given casted votes, is placed on the wrong side of the opinion boundary. Figs 5-6 show that WI is in fact highly underrepresented in the data while opinion variation is moderately large, which fully accounts for the discrepancy between the model and actual votes. Research on Wisconsin's voting during the 2016 elections has shown that Wisconsin was the state with the highest percentage of voters who finalized their voting decision in the final week and voted for Trump (59%, [51]). We argue that this fact, together with the so-called "Shy Trump effect" (Trump supporter were unwilling to reveal their true preference because their support was socially undesirable), has led to the underrepresentation of opinions among Trump supporters in social media. In sum, the best practice is to examine the representativeness of opinion measures and the opinion variation together (combining Figs 5 and 6) in order to evaluate the usefulness of the constructed measure and to unveil the underlying opinion patterns.

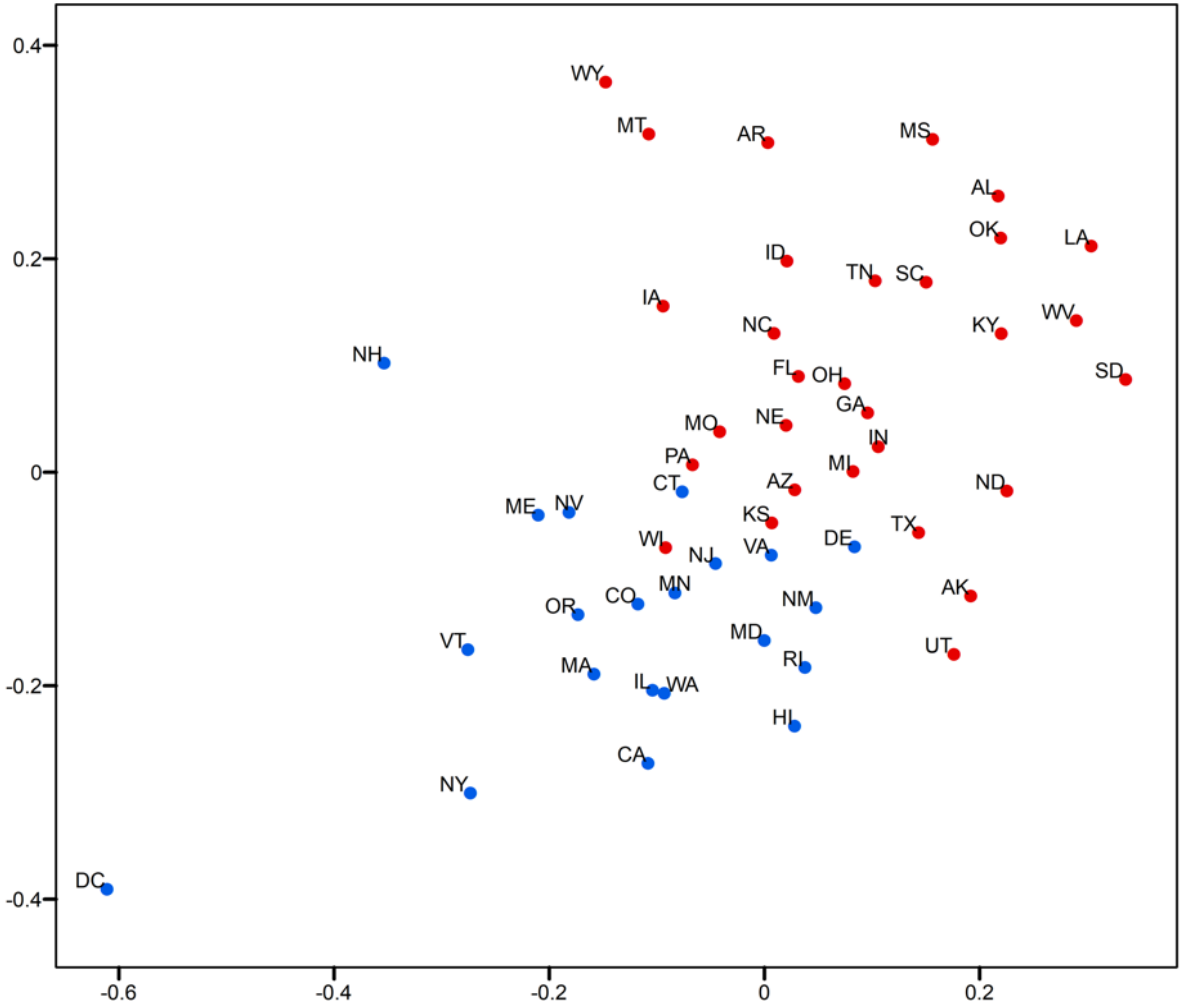

**Fig 3. Two-dimensional MDS representation of the relative opinion space color-coded by the 2016 US presidential popular vote results.** Red: vote for Trump; blue: vote for Clinton.

**Fig 4. Two-dimensional MDS representation of the relative opinion space color-coded by the 2016 US presidential popular vote ratios.** Darker blue means lower ratio of votes for Trump to votes for Clinton; Darker red means higher ratio of votes for Trump to votes for Clinton. Light blue and red depict swing states.

**Fig 5. Two-dimensional MDS representation of the relative opinion space with opinion variation for each state and color-coded by the 2016 US presidential election results.** Red means vote for Trump; Blue means vote for Clinton. A larger circle means larger standard deviation of users' opinion positions for a state.

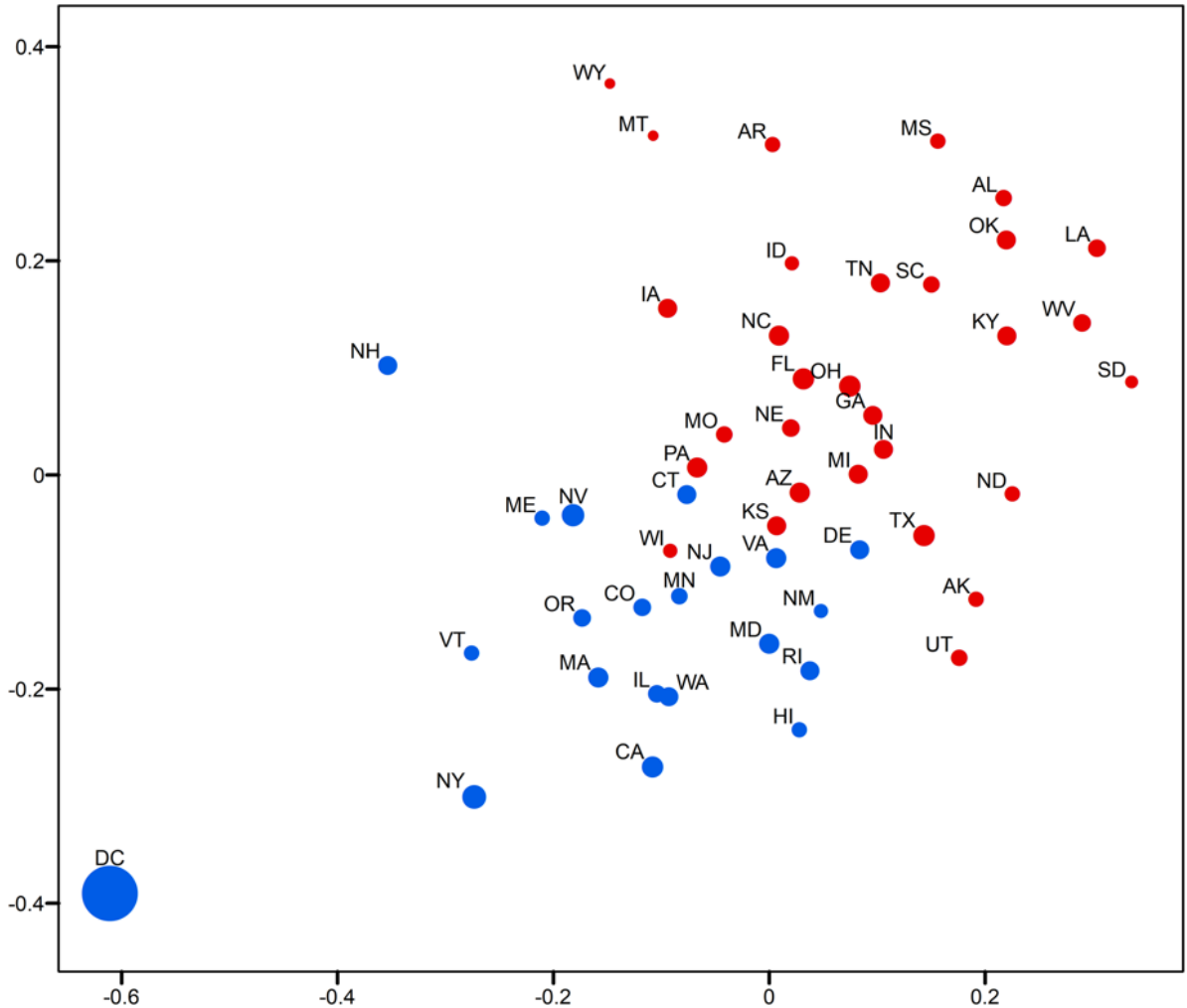

**Fig 6. Two-dimensional MDS representation of the relative opinion space with the ratio of number of twitter users captured in the data to the size of population for each state and color-coded by the 2016 US presidential election results.** Red means vote for Trump; Blue means vote for Clinton. A larger circle means larger ratio.

## Prediction with state-level relative opinion measures

This section showcases the power of the relative opinion measure in supporting opinion predictions at the state level. To achieve this, the continuous representation of the relative opinion needs to be discretized into opinion categories, which can be validated against the real election outcomes and compared to opinion polls at the state level. To transform the continuous opinion to discrete ones, a label propagation method is devised here by taking advantage of the graph structure embedded in the relative opinion measure. Specifically, relative opinion embedding can be used to construct a graph, where each data point (state) is a vertex labelled to denote one of the discrete opinion categories and where an edge exists between a pair of data points based on their proximity in the relative opinion space. Hence, predicting the unknown opinion labels in the graph from a few data points with known labels (prior information) can be formulated as a semi-supervised label propagation problem [52,53]. In the case of the US presidential election, a few "deep-red" and "deep-blue" states whose voters predominantly choose either the Republican (red) or the Democratic (blue) candidate are usually easy to identify through historical voting, which provides the prior information for known labels and makes the above prediction problem feasible.

## Linear neighborhood propagation

We adopt here a well-established label propagation method named Linear Neighborhood Propagation (LNP; [52,53]) to discretize the relative opinions. As a semi-supervised learning approach to propagating labels on a graph, LNP considers the global structure of the graph by capturing the local structure or neighborhood for each vertex (data point) comprising other vertices in close proximity to it [54]. Inspired by some of the nonlinear dimensionality reduction methods, constructing a low dimensional representation of high-dimensionality data with the local structure of the data preserved, such as Locally Linear Embedding (LLE, [55]), LNP assumes the data points are sampled from an underlying manifold and each data point and its label can be linearly reconstructed from its neighbors. That is, there exists an adjacency weight matrix $W$ that minimizes the reconstruction error $\varepsilon$ for $n$ data points:

$$\begin{cases} \min \varepsilon(W) = \sum_{i=1}^{n} \left| x_i - \sum_{j=1}^{k} w_{ij} x_i \right|^2 \\ \qquad s.t. \ \sum_{j=1}^{k} w_{ij} = 1 \end{cases} \quad (2)$$

where $k$ is the number of neighbors for each data point.

Because of the local linearity assumption, the manifold can be well approximated by a sufficiently small neighborhood (a linear subspace or the tangent space) surrounding any data point, which would ideally shrink $\varepsilon$ to zero. The $W$ to be determined essentially characterizes the linear neighborhood for every data point by specifying the contribution of each neighbor. Because $W$ captures the intrinsic local structures of the manifold, the weights are invariant to linear transformations of the high-dimensionality manifold into a low-dimensionality representation preserving the intrinsic structures of the original manifold. This is the fundamental rationale behind LLE. Along a similar rationale, the label $l_i$ of each data point can be reconstructed by a linear combination of its neighbors' labels:

$$l_i = \sum_{j=1}^{k} w_{ij} l_j \qquad (3)$$

The algorithm applied in this study (Algorithm 1) is an extended version of the original algorithm proposed in [52]. Given a set of n data points $x_i$ (aggregate opinion embedding), each has an initial label $l_i$, a vector of length $C$, where each element represents one of the $C$ categories. For a known label $l_i$, one element has the value of 1 indicating the represented category, and all other elements have the value of 0. For an unknown label $l_i$, all elements are initialized with the value of 0. The algorithm includes five general steps as follows. 1) Determine the local linear neighborhood for each data point $x_i$ by computing its $k$ nearest neighbors based on a chosen distance metric (e.g., Euclidean or Geodesic distance); 2) Construct a directed graph $G_{knn}$ by creating an edge from $x_i$ to each of its neighbors; 3) Minimize Equation 2 to calculate $w_{ij}$ for each edge in $G_{knn}$ characterizing local neighborhoods; 4) Apply $w_{ij}$ to Equation 3 to compute label $l_i$ for each $x_i$ in $G_{knn}$ and iterate this process until every label $l_i$ in $G_{knn}$ converges; 5) Determine the category of label $l_i$ by setting the largest element in $l_i$ as 1 and the remaining elements as 0.

Research has shown that the Geodesic distance is superior to the default setting of Euclidean distance in LLE, and can eliminate the "short circuit" problem and lead to a more faithful representation of the global structure of the underlying manifold [56]. The Geodesic distance is approximated as the length of shortest path between a pair of data points in a weighted graph $G_{geo}$, which can be computed as in [57]. Following [56], $G_{geo}$ is constructed by connecting each data point with a set of neighboring data points based on a typical dissimilarity measure, e.g., the Euclidean distance, as the edge weight. To determine the neighboring data points, a minimum value of the neighborhood size is chosen such that all the pairwise geodesic distances are finite [56]. In Step (1.5), an embedding $X' \in \mathbb{R}^d$ is reconstructed from the pairwise geodesic distances via MDS with the same dimensionality as of $X$. Thus, $X'$ can be viewed as an unfolding of the original manifold represented by $X$ to enforce linearity in the reconstructed geometry where the topology of the original manifold is preserved by Geodesic distance. This step ensures the fulfilment of the assumptions by LNP.

---

**Algorithm 1: Linear Neighborhood Propagation**

**Input**:
Data points $X = \{x_1, x_2, \ldots, x_m, x_{m+1}, \ldots, x_n\} \in \mathbb{R}^d$, $\{x_i\}_{i=1}^m$ are labeled, $\{x_i\}_{i=m+1}^n$ are unlabeled.
The initial labels $= \{l_i\}_{i=1}^n$.
The number of nearest neighbors $k$.
**Output**:
The labels for all data points.
**Procedure**:
(1) Compute $k$ nearest neighbors for each data point in $X$ based on a defined distance metric (Euclidean or Geodesic distances); if Euclidean distance is used, skip (1.5), otherwise perform (1.5).
(1.5) Run MDS algorithm on pairwise Geodesic distances to reconstruct $X' \in \mathbb{R}^d$ as an unfolding of $X$ and set $X = X'$.
(2) Construct the $k$-nearest-neighbor graph $G_{knn}$.
(3) Compute the adjacency weight matrix $W$ that best reconstructs each data point in $X$ from its $k$ nearest neighbors by minimizing Equation 2.
(4) For $i = m$ to $n$, Do
　　set $l'_i = \sum_{j=1}^k w_{ij} l_j$ based on Equation 3
　　update $l_i = l'_i$ until $l_i$ converges
(5) For $i = m$ to $n$, Do

> set $l_i = f(l_i)$, where $f(\cdot)$ takes the vector $l_i$ and set its largest element as 1 and any other element as 0.

## Comparison of predictions with Euclidean and Geodesic distances

Given the embedding of data points and the initial labels, the only parameter in Algorithm 1 is the number of nearest neighbors $k$. The influence of $k$ on the quality of embedding generated by LLE and its variants has been studied [55,58]. General criteria must be considered to set the range of $k$. First, the dimensionality of the output embedding should be strictly less than $k$; second, a large $k$ will violate the assumption of local linearity in the neighborhood for curved data sets and lead to the loss of nonlinearity for the mapping. In the following experiments aimed at comparing Algorithm 1 implemented with Euclidean and Geodesic distances, respectively, sensitivity analysis is conducted on $k$ in the range of $[2, \frac{d}{2}]$, where $d$ is the dimensionality of the original embedding. For prediction experiments, sensitivity analysis is also conducted by varying numbers of initial labels that are balanced (2, 4, 6, and 8 labelled states for each of the two opinion classes, i.e., 4, 8, 12, and 16 labels in total) and randomly selected from the set of true labels (the 2016 U.S. presidential popular vote results by state).

According to Equation 2, given the embedding $X$ based on either Euclidean or Geodesic distances and given a $k$ value, an optimal $W$ and the low-dimensionality embedding based on $W$ can be deterministically computed. The quality of $W(X, k)$, viewed as a function of $X$ and $k$, can be evaluated by measures comparing original data points and the embedding mapped into the low-dimensionality space. According to Equation 3, a high-quality $W$ would lead to a good prediction. Two such measures [56], namely Neighborhood Preservation (NP, Equation 4) and Stress (ST, Equation 5), were proposed to assess the quality of $W$ for LLE and they are used here to evaluate the influence of Euclidean and Geodesic distances on the quality of results in relation to the prediction results by Algorithm 1:

$$NP = \frac{1}{n}\sum_{i=1}^{n} P_k(x_i) \quad (4)$$

where $P_k(x_i)$ is the percentage of the $k$-nearest neighbors of $x_i$ in the original space that are preserved in the low-dimensionality embedding space.

$$ST = \frac{\sum_{ij}[\delta(x_i,x_j)-\tau(x_i,x_j)]^2}{\sum_{ij}\tau(x_i,x_j)^2} \quad (5)$$

where $\delta(x_i, x_j)$ and $\tau(x_i, x_j)$ are the pairwise distance in the original and in the embedding space, respectively. As the MDS method in step 1.5 of Algorithm 1 involves randomness, 50 runs are conducted here to examine the stability of the two measures for Geodesic distance.

Figs 7 and 8 present a comparison of NP and ST across a range of $k$ values in [2, 25] for using Euclidean and Geodesic distances in LLE, respectively. Fig 7 shows that the Geodesic distance leads to better neighborhood preservation with the derived $W$ than Euclidean distance does for all values of $k$, except 25. Similarly, the Geodesic distance achieves lower ST for most $k$ values, except for a few cases (10, 11, 12, 16, and 22). Both measures indicate that the Geodesic distance generates superior quality of $W$ than the Euclidean one does in general and that it should lead to better prediction results for Algorithm 1.

Figs 9-12 show the median prediction error of Algorithm 1 with the Euclidean distance, versus the Geodesic distance, for varying numbers of initial labels. Due to the random assignment of the initial labels, 50 runs for each setting are conducted to examine the stability of performance via a 95% confidence interval (the region surrounding the median). Across figs 9-12, similar patterns are shown for each distance type. For the Euclidean distance (EUC), the prediction error first decreases as $k$ increases, then reaches its lowest value

around $k$=8 before it resumes increasing for larger $k$s. After $k$=8, the general trend of the prediction error is increasing, though it decreases slightly after $k$=20. On the other hand, the prediction error for the Geodesic distance (GEO) first decreases until it arrives at a local minimum around $k$=5; after that it bounces back to a local peak around $k$=8. For $k$=9 and onward, the prediction error becomes smaller, which presents a general decreasing trend for the entire curve. It is noted that although predictions with EUC seem always better than those with GEO locally around $k$=8, the latter shows a superior performance globally over a longer range of $k$=[13, 25]. The superiority achieved while using the Geodesic distance should be attributed to Step (1.5) in Algorithm 1 where the linearity is enforced by MDS in the reconstructed geometry of embedding.

The comparison between Figs 9-12, the results with different numbers of initial labels, indicates that prediction errors for both EUC and GEO generally shift lower and the confidence interval width becomes narrower as the number of initial labels increases. This is especially pronounced for GEO in the range of $k \in$[13, 25]. For instance, the performance for models with 8 initial labels dramatically increases compared to that for models with 4 initial labels, as the median error is within [4, 6] for the former while it is between [8, 13] for the latter. It confirms that predictions with more prior information (larger number of initial labels) will lead to consistently better and more stable performance. As we have demonstrated that predictions with GEO consistently outperform those with EUC and achieve global optima over the range of $k \in$[13, 25] under different parameter settings, it remains to determine whether there will always exist an optimal $k$ or a range where some optimal $k$s reside and whether it is possible to identify them before running predictions.

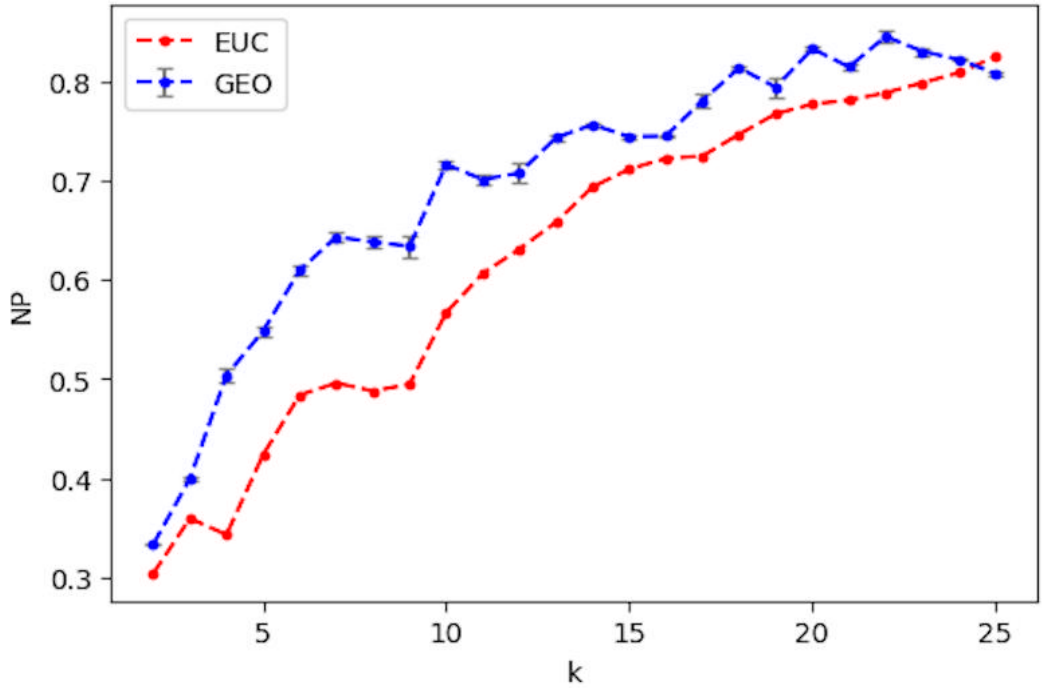

**Fig 7. Comparison of NP when using the Euclidean distance (EUC) and the average NP when using the Geodesic distance (GEO) over a range of *k*.** The grey bar indicates the standard deviation for 50 runs of NP with GEO.

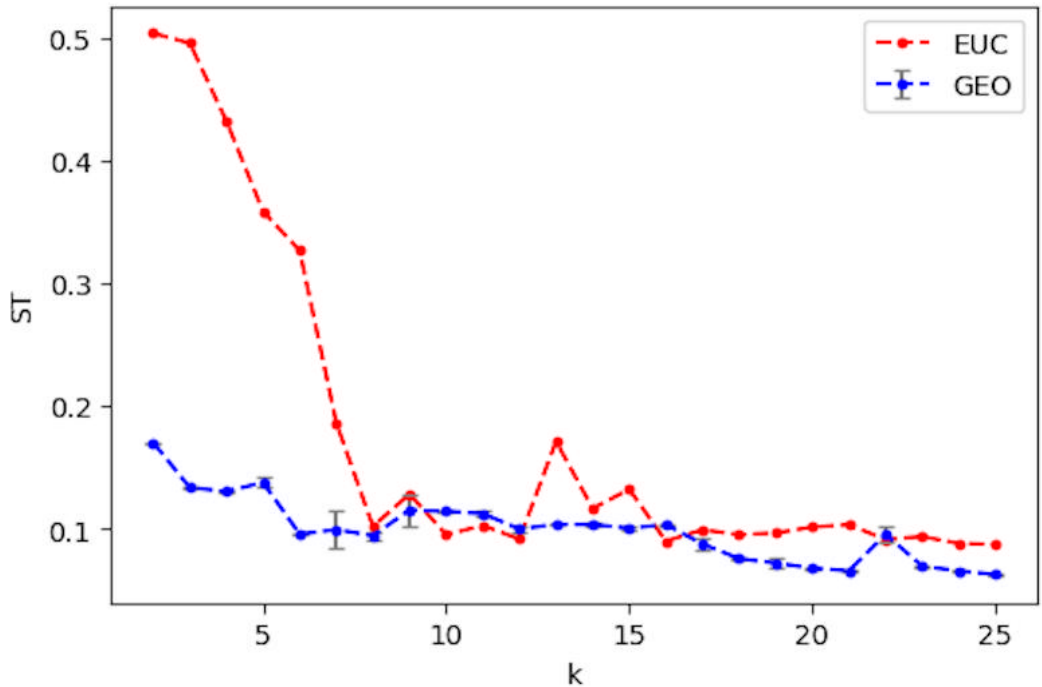

**Fig 8. Comparison of ST when using the Euclidean distance (EUC) and the average ST when using the Geodesic distance (GEO).** The grey bar indicates the standard deviation for 50 runs of ST with GEO.

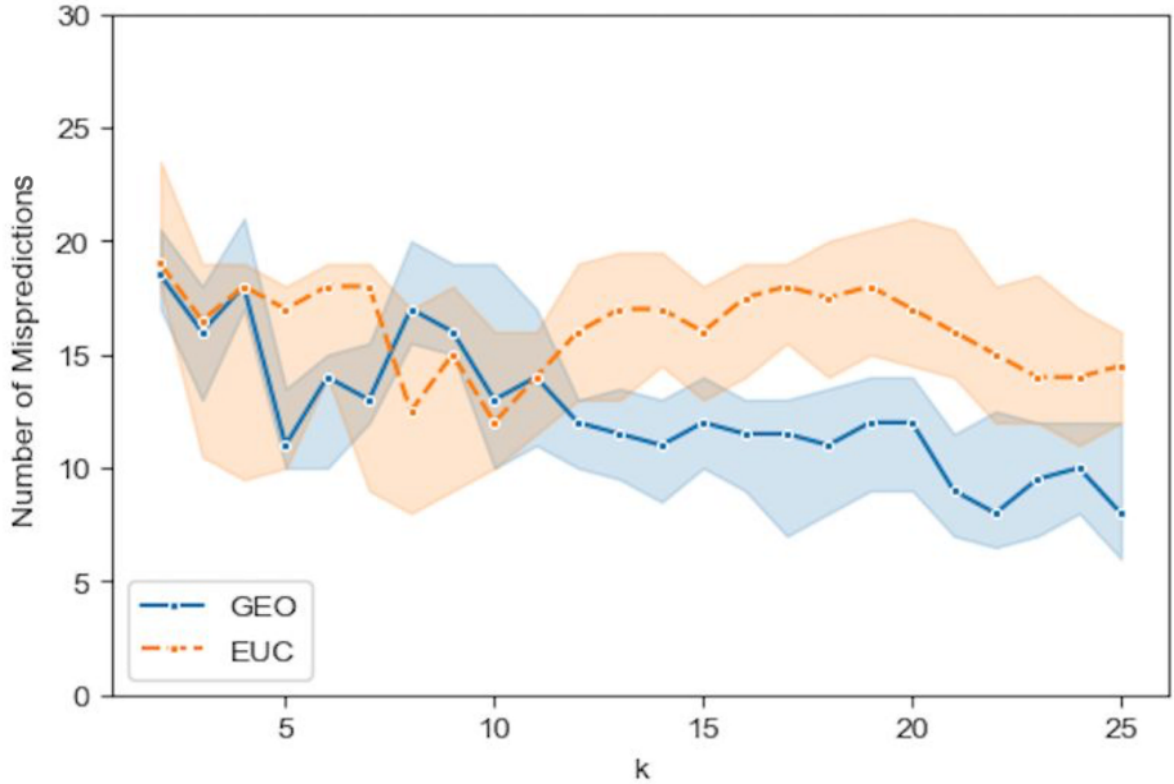

**Fig 9. Comparison of the median prediction errors with the Euclidean distance (EUC) and those with the Geodesic distance (GEO) using 4 initial labels.** The region surrounding each median is a 95% confidence interval calculated from 50 runs of prediction.

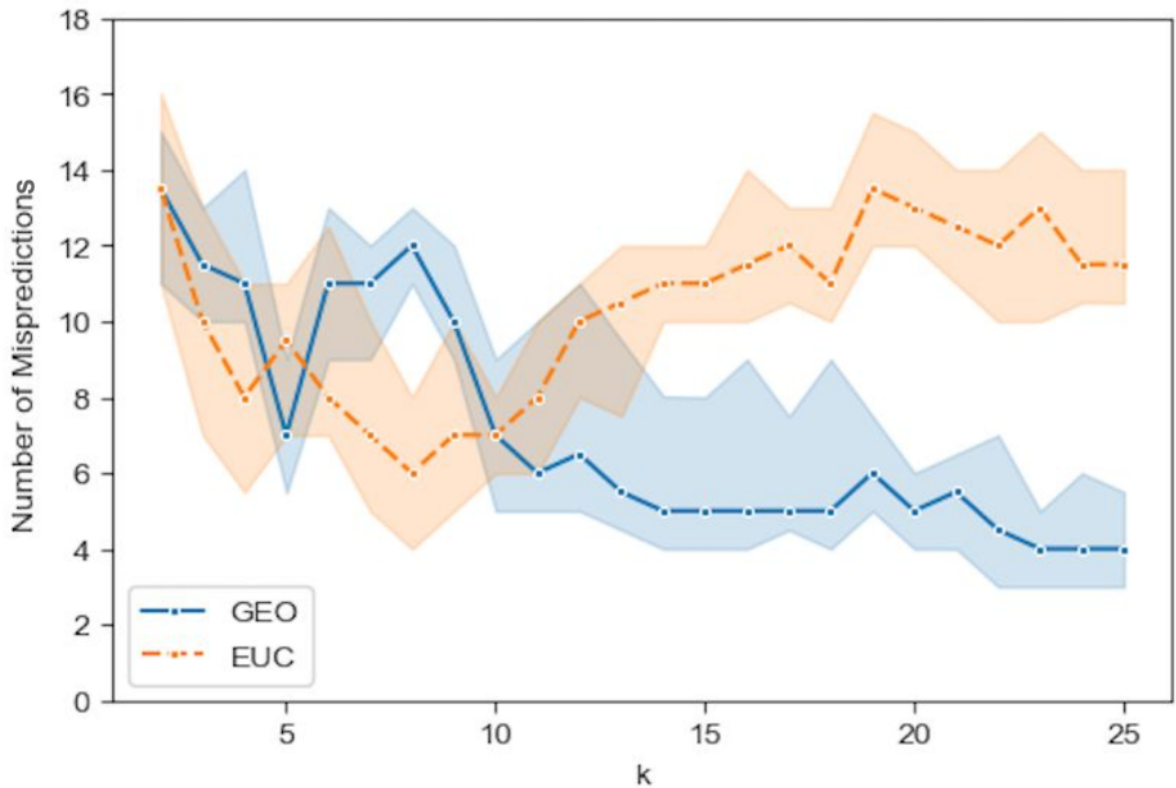

**Fig 10. Comparison of the median prediction errors with the Euclidean distance (EUC) and those with the Geodesic distance (GEO) using 8 initial labels.** The region surrounding each median is a 95% confidence interval calculated from 50 runs of prediction.

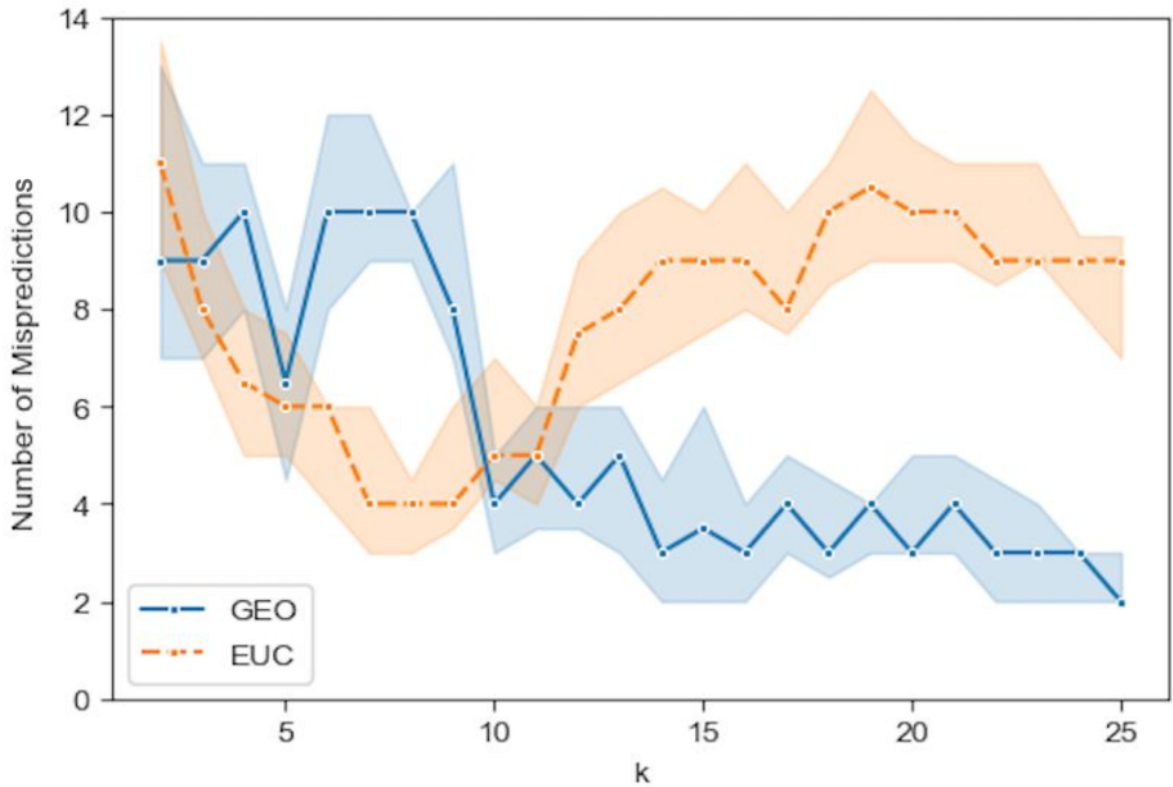

**Fig 11. Comparison of the median prediction errors with the Euclidean distance (EUC) and those with the Geodesic distance (GEO) using 12 initial labels.** The region surrounding each median is a 95% confidence interval calculated from 50 runs of prediction.

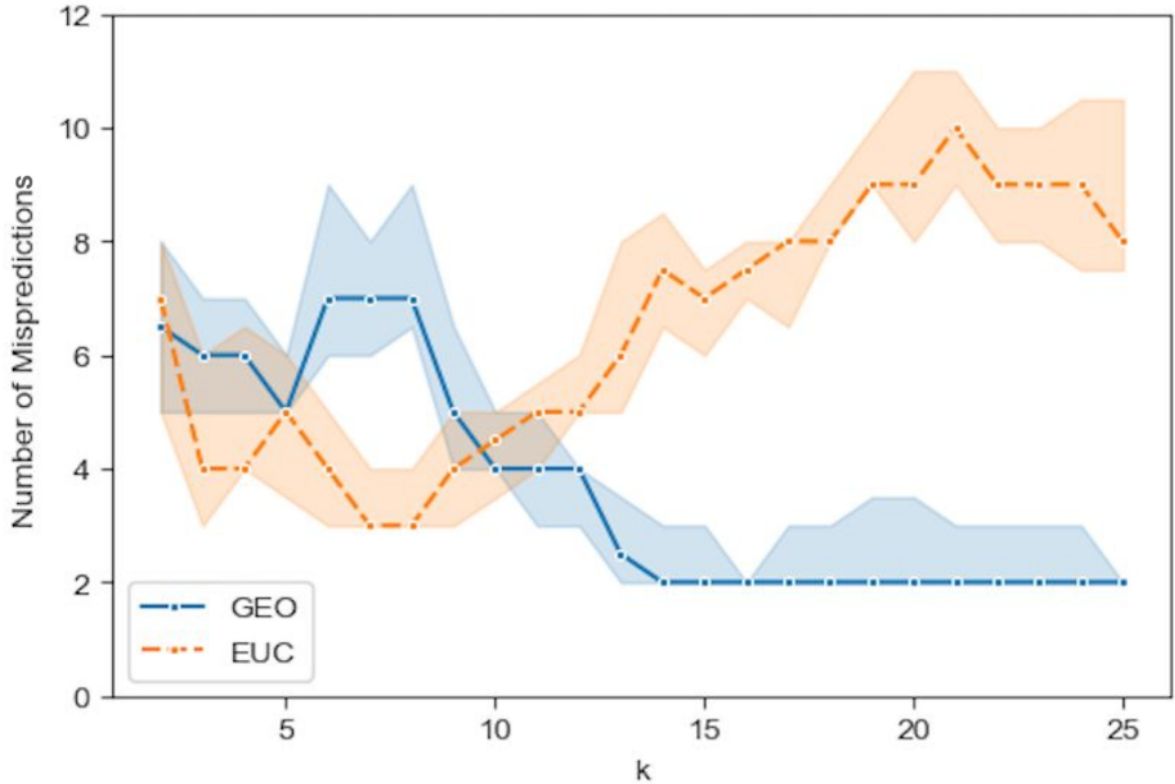

**Fig 12. Comparison of the median prediction errors with the Euclidean distance (EUC) and those with the Geodesic distance (GEO) using 16 initial labels.** The region surrounding each median is a 95% confidence interval calculated from 50 runs of prediction.

## Predictions with the optimal neighborhood sizes

As one of the key parameters in Algorithm 1, the neighborhood size $k$ dramatically affects the quality of prediction as demonstrated in Figs 9-12. To obtain the optimal results, the selection of optimal $k$ becomes a key issue. However, the two measures used in the previous section only enable the comparison between two sets of embedding $X$ based on either Euclidean or Geodesic distances, but they are not suitable for comparison across the values of $k$. As $\boldsymbol{W}(X,k)$ is a function of $k$, an automatic technique, called Preservation Neighborhood Error (PNE), was developed for choosing the optimal $k$ by evaluating the quality of $\boldsymbol{W}$ across a range of $k$ [58]. This technique minimizes a cost function that considers both the local and global geometry preservation:

$$\begin{cases} \boldsymbol{k}_{PNE} = \min_{k} PNE \\ PNE(\boldsymbol{k}) = \frac{1}{2n}\sum_{i=1}^{n}\left[\sum_{j\in K_h}\frac{|\delta(x_i,x_j)-\tau(x_i,x_j)|^2}{k} + \sum_{j\in K_l}\frac{|\delta(x_i,x_j)-\tau(x_i,x_j)|^2}{k}\right] \end{cases} \quad (6)$$

where $K_h$ is the set of $k$ nearest neighbors found in the original space; $K_l$ is the set of $k$ nearest neighbors found in the low-dimensionality embedding space. $\delta(\boldsymbol{x_i}, \boldsymbol{x_j})$ and $\tau(\boldsymbol{x_i}, \boldsymbol{x_j})$ are the pairwise distances in the original and in the embedding space, respectively. In Equation 6, the first item is the error of misses indicating the preservation of local neighborhood, while the second item refers to false positives that reflect the loss of global geometry of the manifold [58].

From the median PNE measure shown in Fig 13 over a range of $k$ with the Geodesic distance, we find that PNE values are lower above $k$=10. Thus, there may exist some optimal $k$s for prediction, especially around $k$=18 and $k$=24. This lies within the [13, 25] range, where the predictions have achieved superior performance in Figs 9-12, although $k$=18, where the minimum of PNE is, does not necessarily correspond to the optimal $k$ for the best prediction. Fig 13 also shows a generally high stability (small confidence intervals) of the PNE measure over the range [13, 25]. It has been demonstrated that PNE is able to indicate a rough range where the optimal neighborhood size $k$ may reside, rather than a specific optimal k, which can significantly reduce the number of prediction runs to select a good model.

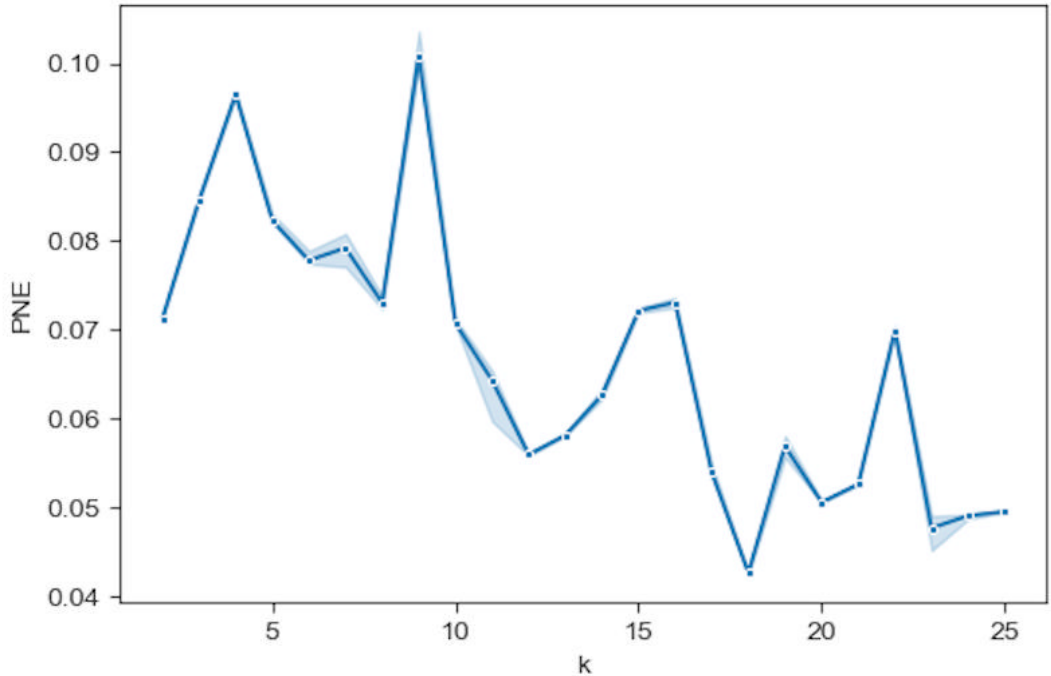

**Fig 13. The median PNE for using the Geodesic distance over a range of neighborhood size $k$.** The region surrounding the median is a 95% confidence interval calculated from 50 runs of PNE.

## Comparison of predictions with polling

This section examines the performance of the opinion predictions with the relative opinion measure in comparison to pre-election polls, which remain the mainstream method to obtain public opinion today. The polling data are from the pre-election wave of the 2016 Cooperative Congressional Election Survey conducted countrywide from October 4th to November 6th [59, 60]. The polling results for Clinton (blue) and Trump (red) are plotted with the relative opinion measure (Fig 14). Compared to Fig 3 showing the actual election votes, the results show 7 misses, including the states of IA, NC, FL, OH, MI, PA, and WI. This performance is better than that for GEO models with 4 initial labels (Fig 9). On the other hand, it is worse than that for GEO models with 8 and more initial labels (Figs 10-12) according to the reported median errors for the suggested range of optimal neighborhood size $k \in [13, 25]$ discussed in the previous section.

Specially, among these misses of polling data there is a cluster of states (IA, NC, FL, and OH) that stand out by their relative opinion measure located a fair distance from the opinion dividing boundary, along which the other misses (MI, PA, and WI) are positioned. It is understandable that the latter cases are difficult to predict, while the former cluster of errors can be corrected by using the relative opinion measure. To demonstrate this, we show here a specific GEO model with 8 predetermined initial labels and $k=18$ chosen from the range of [13, 25]. Eight initial labels are used here because: 1) the prior information to determine 4 deep-red and 4 deep-blue states is trivial given the common knowledge of historical voting; 2) the GEO model with 8 initial labels has shown a good balance of performance and amount of prior information needed. Table 2 details the selection of the 8 initial labels given a general perception of deep-red and deep-blue states (other combinations of deep-red and deep-blue states have been tested, which give similar results). Verification of this model with varying $k$ is performed and the results are shown in Fig 15. It shows a superior and stable model performance over the range of $k \in [17, 24]$ with prediction errors equal to 2. Fig 16 plots the prediction results for all states by this model when $k=18$. The two errors are for the states of WI and KS, which are along the opinion dividing boundary and admittedly difficult to predict. By contrast, the cluster of errors for IA, NC, FL, and OH obtained based on polling do not surface here because, in the relative opinion space, these states are relatively close to the Support-Trump opinion extreme where most deep-red states are located. It is thanks to the nature of the relative opinion measure triangulating every state's opinion position based on its relationship with every other state's position that reduction of uncertainty is achieved and that a measurement of opinions more robust than polling is produced.

Interestingly, when additional labels are assigned to the states of DE, CT, KS, and WI as part of the LNP learning process, the GEO model with 12 initial labels (specified in Table 2) produces only one or no error over the range $k \in [16, 24]$ (Fig 15), which emphasizes the criticality of the prior knowledge about the opinions of states located along and close to the opinion dividing boundary. This prior information could be obtained by accurate opinion polls. Combining the critical prior information with the relative opinion measure would lead to a high quality of predictions that is beyond reach by either of the two alone. In other words, opinion poll can complement the relative opinion measure by providing prior knowledge for initial labels.

**Table 2 Settings for initial labels**

| Class | 8 Labels | 12 Labels |
|---|---|---|
| **Clinton (blue)** | CA, DC, MA, NY | CA, DC, MA, NY, DE, CT |
| **Trump (red)** | NE, OK, WV, WY | NE, OK, WV, WY, KS, WI |

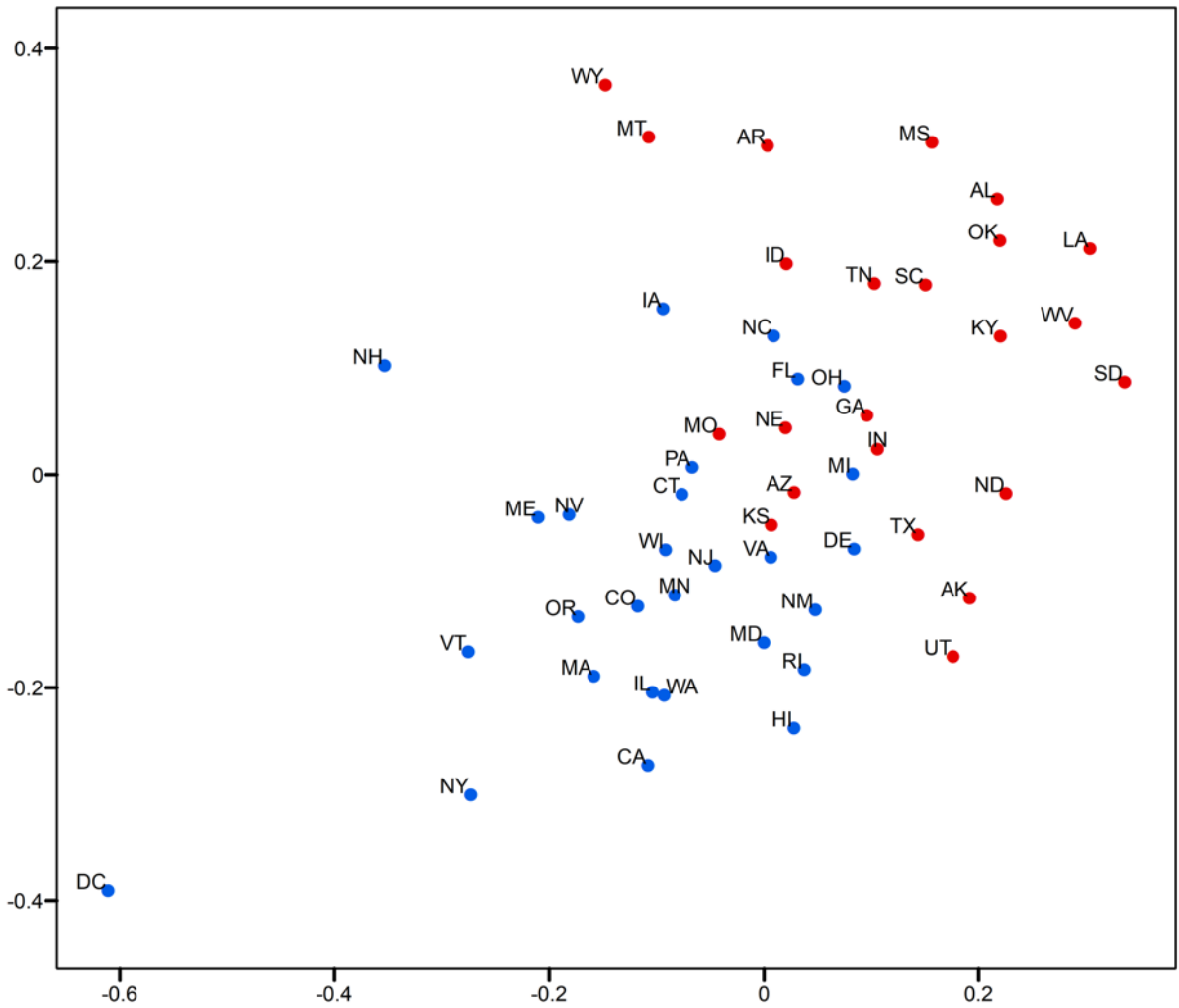

**Fig 14. Two-dimensional MDS representation of the relative opinion space with 2016 Cooperative Congressional Election Survey polling data (conducted from October 4th to November 6th, 2016).** Red: vote for Trump; blue vote for Clinton.

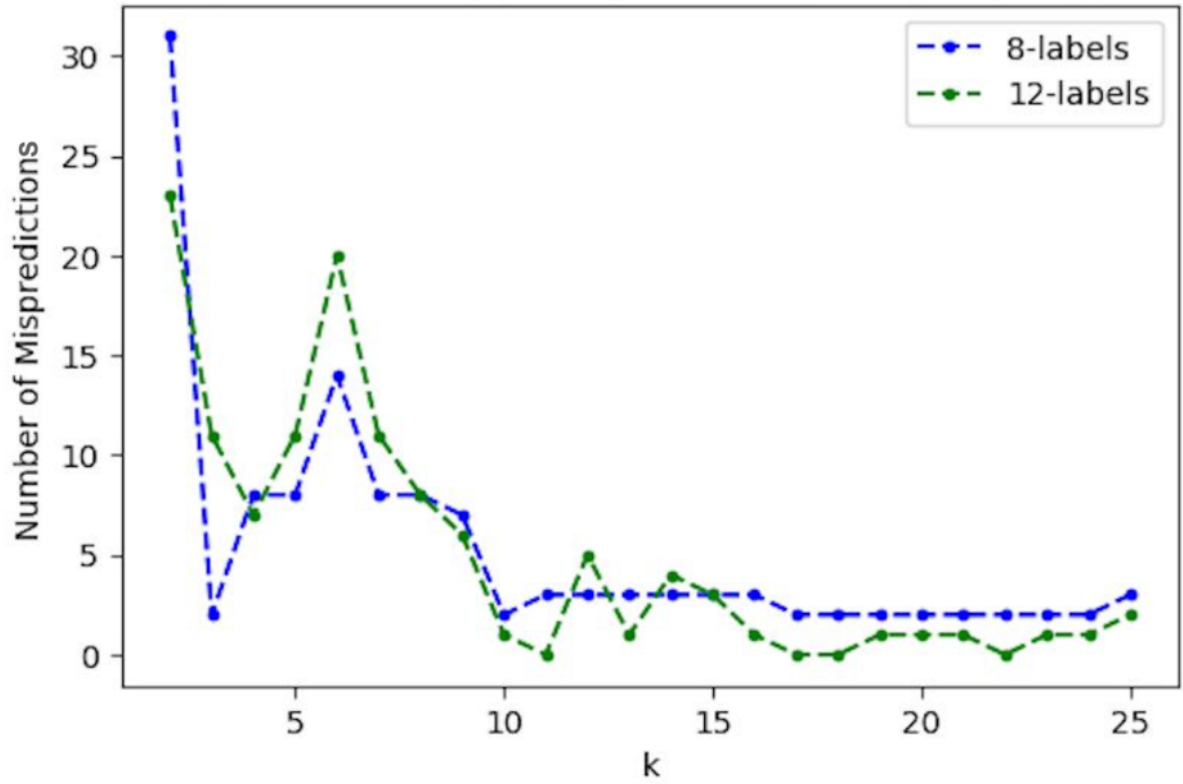

**Fig 15. Prediction errors for the two models with 8 and 12 initial labels, respectively, using settings in Table 2 and the Geodesic distance**.

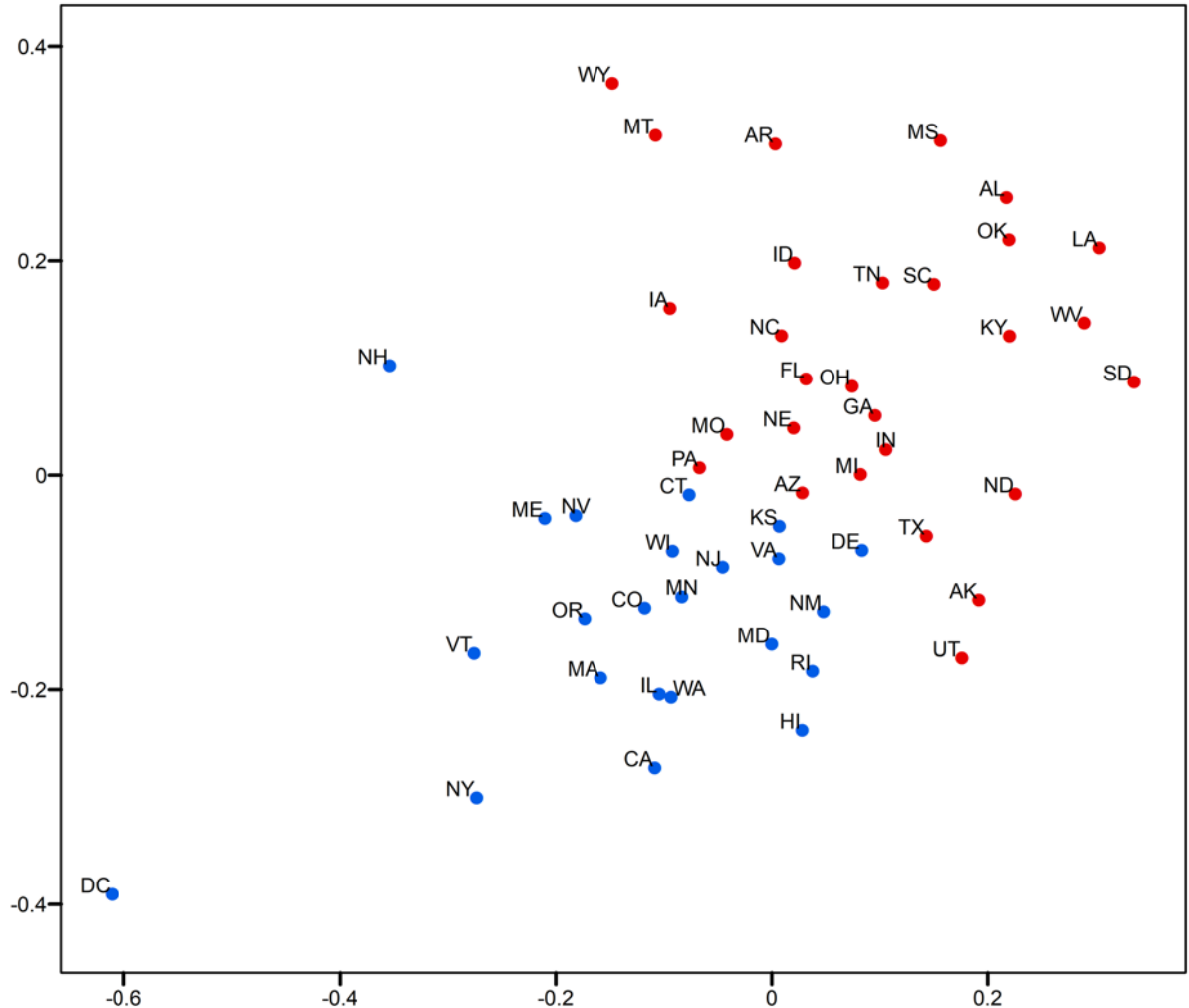

**Fig 16. Two-dimensional MDS representation of the relative opinion space with prediction results of the model with 8 initial labels and neighborhood size *k*=18.**

# Conclusions and future work

This study proposed to measure relative opinion from LBSM data in response to the challenge of leveraging the rich and unstructured discourse on social media as an alternative source to opinion polls for public opinion measurement (the first question in the introduction). The advantages of the relative opinion measure lie in its theoretical grounding and methodological suitability to LBSM data. The relative opinion conceptualization theoretically compensates the deficiency of the absolute opinion measure in representing complex opinion structures. On the other hand, the pairwise relationship of opinions characterized by this measure naturally suits the embedding representation of opinion positions that can be learned from high-dimensionality textual messages with supervision. To make this quantification technically feasible, a modeling framework was proposed, including building a training dataset by adopting a state-of-the-art approach and developing a supervised learning method, the opinion-oriented word embedding.

To demonstrate the validity of the relative opinion measure, spatial visualizations of relative opinion space were constructed to aid visual analytics. As an exploratory analysis approach, it facilitates the examination of uncertainty and representativeness of the measure, the discovery of opinion patterns across geographies, and the correspondence between relative opinion positions with other variables such as opinion polls and real election outcomes, which may stimulate the formation of new hypotheses on electoral behavior. Furthermore, the relative opinion measure supports practical opinion predictions at aggregated geographic levels, transforming a continuous representation into a discrete one that is comparable to opinion polls and strongly validated by election outcomes. This is enabled by a linear neighborhood propagation method that incorporates the intrinsic geometry of the relative opinion space, optimal neighborhood sizes, and the prior knowledge of opinion preferences for a small number of entities.

In the case study of the 2016 U.S. presidential election, it is demonstrated that the relative opinion measure constructed on Twitter data is more robust than polling data, thanks to its theoretical grounding and the modeling framework that exploits the intrinsic properties of LBSM data. However, given their differences in concept, data collection, and methodology, the relative opinion measure cannot and should not replace polling. Instead, the two type of measures and their associated data are complementary in opinion measurement, which has been demonstrated by our prediction approach to be feasible and promising. This is an answer to the second research question presented in the introduction.

Admittedly, as the present work is an initial study of the relative opinion measure, further investigation is needed. There are several directions for future studies. First, the results reported in this study are based on the 1% sample of tweets for the study period. If this sampling rate can increase to better represent the population, we should be able to examine the sensitivity of our measure to the variation of sample size. Second, measuring relative opinion at a finer geographic level, such as county or city, could be studied with tweets geolocated by enhanced spatial resolution; clustering opinions can be done to evaluate how the boundaries of opinion groups are different from geographic boundaries [61]. Third, with tweets extracted longitudinally, the temporal dynamics of this measure could be investigated to support opinion predictions over time. Last, social network data from Twitter could be utilized and incorporated into the relative opinion measure for better opinion measurement.

## Acknowledgments

We acknowledge the insightful discussions with Professor Cherie Maestas and the polling data source she suggested.

## References

1. Bohannon J. The pulse of the people. Science. 2017;355: 470–472. doi:10.1126/science.355.6324.470

2. Klašnja M, Barberá P, Beauchamp N, Nagler J, Tucker JA. Measuring Public Opinion with Social Media Data. The Oxford Handbook of Polling and Survey Methods. 2018. Available: https://www.oxfordhandbooks.com/view/10.1093/oxfordhb/9780190213299.001.0001/oxfordhb-9780190213299-e-3

3. Bovet A, Morone F, Makse HA. Validation of Twitter opinion trends with national polling aggregates: Hillary Clinton vs Donald Trump. Sci Rep. 2018;8: 8673. doi:10.1038/s41598-018-26951-y

4. Murphy J, Link MW, Childs JH, Tesfaye CL, Dean E, Stern M, et al. Social Media in Public Opinion ResearchExecutive Summary of the Aapor Task Force on Emerging Technologies in Public Opinion Research. Public Opin Q. 2014;78: 788–794. doi:10.1093/poq/nfu053

5. Berinsky AJ. Measuring public opinion with surveys. Annu Rev Polit Sci. 2017;20: 309–329. doi:10.1146/annurev-polisci-101513-113724

6. Zaller JR. The nature and origins of mass opinion. Cambridge university press; 1992.

7. Tourangeau R, Rips LJ, Rasinski K. The psychology of survey response. Cambridge University Press; 2000.

8. Zaller J, Feldman S. A Simple Theory of the Survey Response: Answering Questions versus Revealing Preferences. Am J Polit Sci. 1992;36: 579–616. doi:10.2307/2111583

9. Beauchamp N. Predicting and interpolating state-level polls using twitter textual data: predicting polls with twitter. Am J Polit Sci. 2017;61: 490–503. doi:10.1111/ajps.12274

10. Heredia B, Prusa JD, Khoshgoftaar TM. Social media for polling and predicting United States election outcome. Soc Netw Anal Min. 2018;8: 48. doi:10.1007/s13278-018-0525-y

11. King G, Lam P, Roberts ME. Computer-Assisted Keyword and Document Set Discovery from Unstructured Text. Am J Polit Sci. 2017;61: 971–988.

12. Fang A, Ounis I, Habel P, Macdonald C, Limsopatham N. Topic-centric Classification of Twitter User's Political Orientation. Proceedings of the 38th International ACM SIGIR Conference on Research and Development in Information Retrieval. New York, NY, USA: ACM; 2015. pp. 791–794. doi:10.1145/2766462.2767833

13. DiGrazia J, McKelvey K, Bollen J, Rojas F. More Tweets, More Votes: Social Media as a Quantitative Indicator of Political Behavior. PLOS ONE. 2013;8: e79449. doi:10.1371/journal.pone.0079449

14. Gayo-Avello D. A Meta-Analysis of State-of-the-Art Electoral Prediction From Twitter Data. Soc Sci Comput Rev. 2013;31: 649–679. doi:10.1177/0894439313493979

15. Gayo-Avello D. Don'T Turn Social Media into Another "Literary Digest" Poll. Commun ACM. 2011;54: 121–128. doi:10.1145/2001269.2001297

16. Jungherr A, Schoen H, Posegga O, Jürgens P. Digital Trace Data in the Study of Public Opinion: An Indicator of Attention Toward Politics Rather Than Political Support. Soc Sci Comput Rev. 2017;35: 336–356. doi:10.1177/0894439316631043

17. Jungherr A. Twitter in Politics: A Comprehensive Literature Review. Rochester, NY: Social Science Research Network; 2014 Feb. Report No.: ID 2402443. Available: https://papers.ssrn.com/abstract=2402443

18. Anuta D, Churchin J, Luo J. Election Bias: Comparing Polls and Twitter in the 2016 U.S. Election. ArXiv170106232 Cs. 2017 [cited 8 Oct 2019]. Available: http://arxiv.org/abs/1701.06232

19. Celli F, Stepanov E, Poesio M, Riccardi G. Predicting Brexit: Classifying Agreement is Better than Sentiment and Pollsters. Proceedings of the Workshop on Computational Modeling of People's Opinions, Personality, and Emotions in Social Media (PEOPLES). Osaka, Japan: The COLING 2016 Organizing Committee; 2016. pp. 110–118.

20. Grčar M, Cherepnalkoski D, Mozetič I, Kralj Novak P. Stance and influence of Twitter users regarding the Brexit referendum. Comput Soc Netw. 2017;4: 6. doi:10.1186/s40649-017-0042-6

21. Grimmer J, Stewart BM. Text as Data: The Promise and Pitfalls of Automatic Content Analysis Methods for Political Texts. Polit Anal. 2013;21: 267–297. doi:10.1093/pan/mps028

22. Ceron A, Curini L, Iacus SM, Porro G. Every tweet counts? How sentiment analysis of social media can improve our knowledge of citizens' political preferences with an application to Italy and France. New Media Soc. 2014;16: 340–358.

23. Pang B, Lee L. Opinion Mining and Sentiment Analysis. Found Trends® Inf Retr. 2008;2: 1–135. doi:10.1561/1500000011

24. Bengio Y. Deep Learning of Representations: Looking Forward. In: Dediu A-H, Martín-Vide C, Mitkov R, Truthe B, editors. Statistical Language and Speech Processing. Springer Berlin Heidelberg; 2013. pp. 1–37.

25. Mikolov T, Sutskever I, Chen K, Corrado GS, Dean J. Distributed Representations of Words and Phrases and their Compositionality. In: Burges CJC, Bottou L, Welling M, Ghahramani Z, Weinberger KQ, editors. Advances in Neural Information Processing Systems 26. Curran Associates, Inc.; 2013. pp. 3111–3119. Available: http://papers.nips.cc/paper/5021-distributed-representations-of-words-and-phrases-and-their-compositionality.pdf

26. Tang D, Wei F, Yang N, Zhou M, Liu T, Qin B. Learning Sentiment-Specific Word Embedding for Twitter Sentiment Classification. Proceedings of the 52nd Annual Meeting of the Association for Computational Linguistics (Volume 1: Long Papers). Baltimore, Maryland: Association for Computational Linguistics; 2014. pp. 1555–1565. doi:10.3115/v1/P14-1146

27. Tang D, Zhang M. Deep Learning in Sentiment Analysis. In: Deng L, Liu Y, editors. Deep Learning in Natural Language Processing. Singapore: Springer Singapore; 2018. pp. 219–253. doi:10.1007/978-981-10-5209-5_8

28. Young T, Hazarika D, Poria S, Cambria E. Recent Trends in Deep Learning Based Natural Language Processing [Review Article]. IEEE Comput Intell Mag. 2018;13: 55–75. doi:10.1109/MCI.2018.2840738

29. Collobert R, Weston J, Bottou L, Karlen M, Kavukcuoglu K, Kuksa P. Natural language processing (almost) from scratch. J Mach Learn Res. 2011;12: 2493–2537.

30. Yang X, Macdonald C, Ounis I. Using word embeddings in Twitter election classification. Inf Retr J. 2018;21: 183–207. doi:10.1007/s10791-017-9319-5

31. Paul D, Li F, Teja MK, Yu X, Frost R. Compass: Spatio Temporal Sentiment Analysis of US Election What Twitter Says! Proceedings of the 23rd ACM SIGKDD International Conference on Knowledge Discovery and Data Mining. New York, NY, USA: ACM; 2017. pp. 1585–1594. doi:10.1145/3097983.3098053

32. Duggan M. Demographics of Social Media Users in 2015. Washington, DC: Pew Research Center; 2015 Aug. Available: https://www.pewinternet.org/2015/08/19/the-demographics-of-social-media-users/

33. Mislove A, Lehmann S, Ahn Y-Y, Onnela J-P, Rosenquist JN. Understanding the Demographics of Twitter Users. Fifth International AAAI Conference on Weblogs and Social Media. 2011. Available: https://www.aaai.org/ocs/index.php/ICWSM/ICWSM11/paper/view/2816

34. Malik MM, Lamba H, Nakos C, Pfeffer J. Population Bias in Geotagged Tweets. Ninth International AAAI Conference on Web and Social Media. 2015. Available: https://www.aaai.org/ocs/index.php/ICWSM/ICWSM15/paper/view/10662

35. Hampton K, Goulet LS, Purcell K. Social networking sites and our lives. In: Pew Research Center: Internet, Science & Tech [Internet]. 16 Jun 2011 [cited 4 Oct 2019]. Available: https://www.pewinternet.org/2011/06/16/social-networking-sites-and-our-lives/

36. Vaccari C, Valeriani A, Barberá P, Bonneau R, Jost JT, Nagler J, et al. Social media and political communication: A survey of Twitter users during the 2013 Italian general election. Riv Ital Sci Polit. 2013. doi:10.1426/75245

37. Diaz F, Gamon M, Hofman JM, Kıcıman E, Rothschild D. Online and Social Media Data As an Imperfect Continuous Panel Survey. PLOS ONE. 2016;11: e0145406. doi:10.1371/journal.pone.0145406

38. Zheng X, Zeng Z, Chen Z, Yu Y, Rong C. Detecting spammers on social networks. Neurocomputing. 2015;159: 27–34. doi:10.1016/j.neucom.2015.02.047

39. Barberá P, Rivero G. Understanding the political representativeness of twitter users. Soc Sci Comput Rev. 2015;33: 712–729. doi:10.1177/0894439314558836

40. Leetaru K, Wang S, Cao G, Padmanabhan A, Shook E. Mapping the global Twitter heartbeat: The geography of Twitter. First Monday. 2013;18. doi:10.5210/fm.v18i5.4366

41. Arthur R, Williams HTP. Scaling laws in geo-located Twitter data. PLOS ONE. 2019;14: e0218454. doi:10.1371/journal.pone.0218454

42. Hopkins DJ, King G. A Method of Automated Nonparametric Content Analysis for Social Science. Am J Polit Sci. 2010;54: 229–247. doi:10.1111/j.1540-5907.2009.00428.x

43. Thill J-C. Is Spatial Really That Special? A Tale of Spaces. In: Popovich VV, Claramunt C, Devogele T, Schrenk M, Korolenko K, editors. Information Fusion and Geographic Information Systems: Towards the Digital Ocean. Berlin, Heidelberg: Springer Berlin Heidelberg; 2011. pp. 3–12. doi:10.1007/978-3-642-19766-6_1

44. Couclelis H. Location, place, region, and space. Geogr Inn Worlds. 1992;2: 15–233.

45. Gimpel K, Schneider N, O'Connor B, Das D, Mills D, Eisenstein J, et al. Part-of-speech Tagging for Twitter: Annotation, Features, and Experiments. Proceedings of the 49th Annual Meeting of the Association for Computational Linguistics: Human Language Technologies: Short Papers - Volume 2. Stroudsburg, PA, USA: Association for Computational Linguistics; 2011. pp. 42–47. Available: http://dl.acm.org/citation.cfm?id=2002736.2002747

46. Duchi J, Hazan E, Singer Y. Adaptive Subgradient Methods for Online Learning and Stochastic Optimization. J Mach Learn Res. 2011;12: 2121–2159.

47. Le Q, Mikolov T. Distributed Representations of Sentences and Documents. Proceedings of the 31st International Conference on International Conference on Machine Learning - Volume 32. JMLR.org; 2014. pp. II–1188–II–1196. Available: http://dl.acm.org/citation.cfm?id=3044805.3045025

48. Borg I, Groenen PJF. Modern multidimensional scaling: theory and applications. 2nd ed. New York: Springer; 2005.

49. Galbraith JK. How income inequality can make or break presidential elections. In: PBS NewsHour [Internet]. 2 Jan 2018 [cited 4 Oct 2019]. Available: https://www.pbs.org/newshour/economy/making-sense/how-income-inequality-can-make-or-break-presidential-elections

50. Lai KKR, Parlapiano A. 42 States Shifted to the Right in 2016. The New York Times. 9 Nov 2016. Available: https://www.nytimes.com/interactive/2016/11/09/us/elections/states-shift.html, https://www.nytimes.com/interactive/2016/11/09/us/elections/states-shift.html. Accessed 4 Oct 2019.

51. Kennedy C, Blumenthal M, Clement S, Clinton JD, Durand C, Franklin C, et al. An Evaluation of the 2016 Election Polls in the United States. Public Opin Q. 2018;82: 1–33. doi:10.1093/poq/nfx047

52. Wang F, Zhang C. Label Propagation through Linear Neighborhoods. IEEE Trans Knowl Data Eng. 2008;20: 55–67. doi:10.1109/TKDE.2007.190672

53. Wang J, Wang F, Zhang C, Shen HC, Quan L. Linear Neighborhood Propagation and Its Applications. IEEE Trans Pattern Anal Mach Intell. 2009;31: 1600–1615. doi:10.1109/TPAMI.2008.216

54. Zhou D, Bousquet O, Lal TN, Weston J, Schölkopf B. Learning with local and global consistency. Advances in neural information processing systems. 2004. pp. 321–328.

55. Roweis ST, Saul LK. Nonlinear Dimensionality Reduction by Locally Linear Embedding. Science. 2000;290: 2323–2326. doi:10.1126/science.290.5500.2323

56. Varini C, Degenhard A, Nattkemper T. ISOLLE: Locally Linear Embedding with Geodesic Distance. In: Jorge AM, Torgo L, Brazdil P, Camacho R, Gama J, editors. Knowledge Discovery in Databases: PKDD 2005. Springer Berlin Heidelberg; 2005. pp. 331–342.

57. Tenenbaum JB, Silva V de, Langford JC. A Global Geometric Framework for Nonlinear Dimensionality Reduction. Science. 2000;290: 2319–2323. doi:10.1126/science.290.5500.2319

58. Álvarez-Meza A, Valencia-Aguirre J, Daza-Santacoloma G, Castellanos-Domínguez G. Global and local choice of the number of nearest neighbors in locally linear embedding. Pattern Recognit Lett. 2011;32: 2171–2177. doi:10.1016/j.patrec.2011.05.011

59. Ansolabehere S, Schaffner BF. Cooperative congressional election study, 2016: Common content. Cambridge, MA: Harvard University; 2017 Aug. Report No.: Release 2.

60. YouGov. Cooperative Congressional Election Study: Clinton leads Trump by 4. In: YouGov [Internet]. 7 Nov 2016 [cited 4 Oct 2019]. Available: https://today.yougov.com/topics/politics/articles-reports/2016/11/07/cces-2016

61. Rodriguez MZ, Comin CH, Casanova D, Bruno OM, Amancio DR, Costa L da F, et al. Clustering algorithms: A comparative approach. PLOS ONE. 2019;14: e0210236. doi:10.1371/journal.pone.0210236

# Supporting information

**S1 Appendix. The procedure of creating training data.**

# S1 Appendix

# A. The procedure of creating training data

The procedure used in the section 'Data collection and training data for opinion identification' is described as follows. It includes five steps:

1) Identifying seed hashtags. Among the most frequent hashtags, we identify four hashtags, each one representing a different opinion category: #maga for pro-Trump (the abbreviation of the official Trump campaign slogan: Make America Great Again), #imwithher for pro-Hillary (the official Clinton campaign slogan), #nevertrump for anti-Trump, and #neverhillary for anti-Clinton.

2) Building co-occurrence networks of hashtags. The hashtag co-occurrence network $H(V, E)$ is then constructed, where the set of vertices $v_i \in V$ represents hashtags, and an edge $e_{ij}$ is drawn between $v_i$ and $v_j$ if they appear together in a tweet. The resulting graph has 108,600 vertices and 881,344 edges.

3) Computing similarity networks of hashtags. With the constructed $H(V, E)$, we test the significance of each edge $e_{ij}$ [1]. Specifically, the probability $p_{ij}$ (p-value of the null hypothesis) to observe the corresponding number of co-occurrences $k$ by chance is computed given the number of occurrences $n_i$ and $n_j$ of the vertices $v_i$ and $v_j$, and the total number of tweets $N$ (Equation 1). Only significant edges satisfying $p_{ij} < p_o$, where $p_o = 10^{-6}$, are kept, effectively filtering out spurious relations between hashtags. Then, a weight $s_{ij}$ is assigned to the edge $e_{ij}$ (based on Equation 2) representing the significance of the similarity between two hashtags. Retaining only significant similarity linkages reduces the graph to 108,600 vertices and 51,633 edges, which comprises a similarity network of hashtags.

$$p_{ij}(k) = \prod_{m=0}^{n_j-k-1} \left(1 - \frac{n_i}{N-m}\right) \prod_{m=0}^{k-1} \frac{(n_i-m)(n_j-m)}{(N-n_j+k-m)(k-m)} \quad (1)$$

$$s_{ij} = \log\left(\frac{p_o}{p_{ij}}\right) \quad (2)$$

4) Classifying hashtags in the similarity network. The similarity network is then used to discover and classify hashtags that are significantly similar to the seed hashtags as in [2,3]. The main idea behind the classification of hashtags is that a vertex $v_i$ determines its label, denoting the class to which it belongs, based on the labels of its neighbors, defined as vertices that have edges with $v_i$. It is assumed that each $v_i$ in the network chooses to obtain the label carried by the largest number of its neighbors, with ties broken uniformly randomly. Formally, if $C_1, \ldots, C_j$ are the labels that are currently active in the network and $d_i^{C_j}$ is the number of neighbors $v_i$ has with label $C_j$, then every $v_i$ keeps acquiring its label until for each $v_i$,

$$\text{If } v_i \text{ has label } C_m, \text{then } d_i^{C_m} \geq d_i^{C_j} \quad (3)$$

5) Network pruning and human validation. We further prune the resulting set of labelled hashtags by retaining only those that meet the following criterion.

$$n_i > r \max_{v_j \in C_m} n_j \quad (4)$$

where $n_i$ is the occurrences of the hashtag associated with vertex $v_i$, $C_m$ is the label of $v_i$ and $r = 0.001$ is a threshold parameter. Finally, discovered hashtags are human validated to select hashtags having direct reference to the candidate, its party or slogans of the candidate and that express an opinion. In practice, the entire process (steps 1-5) is iterative until a stable set of hashtags is found. Table 1 shows the top-10 labelled hashtags with the highest occurrence for each category. The final network results in four different clusters corresponding to the Pro-Clinton, Anti-Clinton, Pro-Trump and Anti-Trump hashtags (Fig 1).

Table 1 Top-10 labelled hashtags with the highest occurrence for each category

| Pro-Clinton | Anti-Clinton | Pro-Trump | Anti-Trump |
|---|---|---|---|
| #imwithher | #dickileaks | #trumppence16 | #dumpthetrump |
| #clintonkaine | #hillaryforprison2016 | #trumprally | #nevertrump |
| #hillary2016 | #neverhillary | #trumpstrong | #lockhimup |
| #votehillary | #crookedhillary | #draintheswamp | #bullytrump |
| #imwithher2016 | #hillarysemails | #maga3x | #nastywomenunite |
| #hillaryforamerica | #lockherup | #maga | #racist |
| #hillaryclintonforpresident | #corrupthillary | #trump2016 | #stoptrump |
| #hillarysarmy | #clintonscandals | #votetrump | #loserdonald |
| #momsdemandhillary | #indicthillary | #makeamericagreatagain | #lyingtrump |
| #womenvote | #notwithher | #americafirst | #anybodybuttrump |

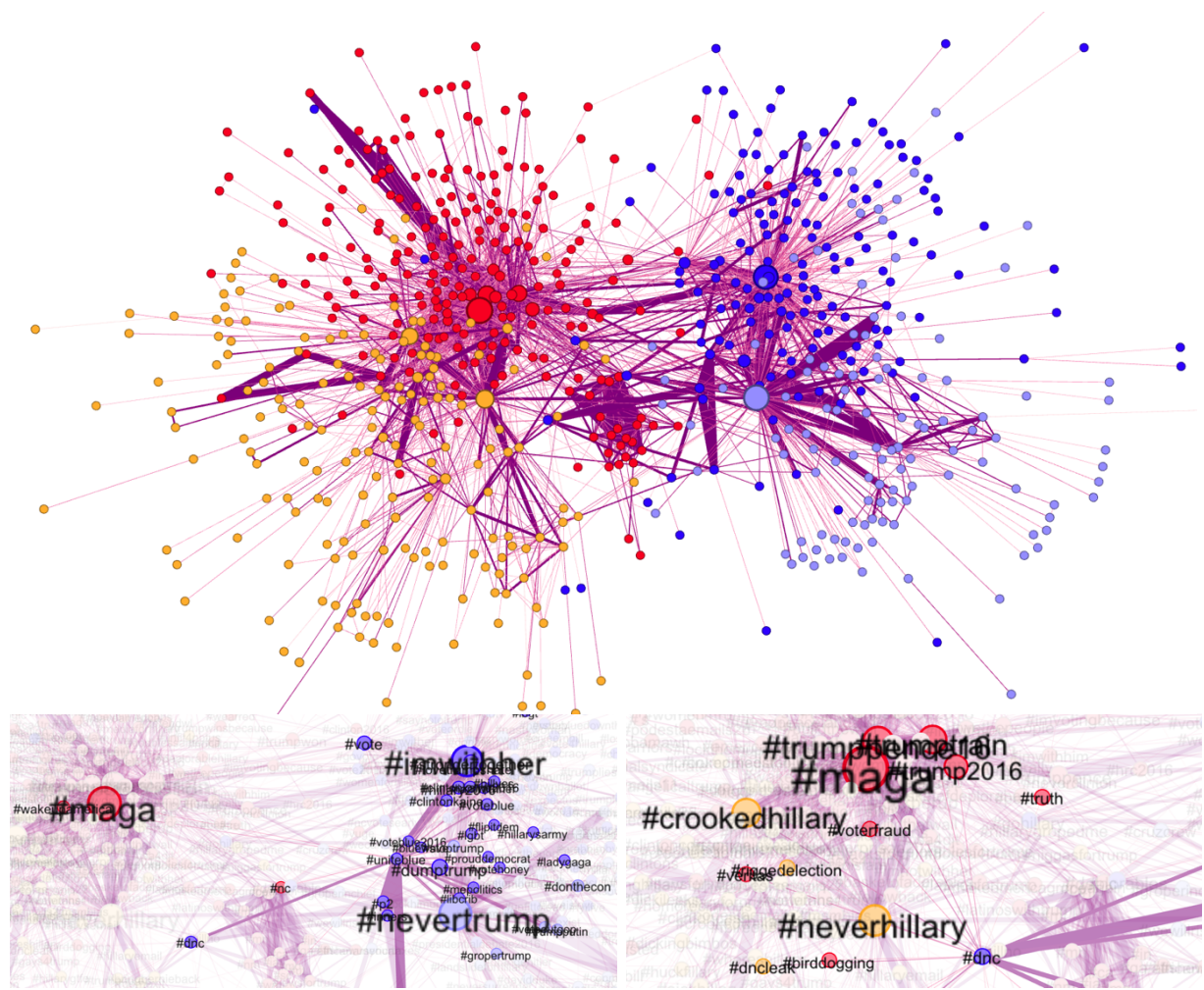

**Fig 1. Clusters of hashtags.** Red: Pro-Trump, Orange: Anti-Clinton, Dark Blue: Pro-Clinton, Light Blue: Anti-Trump.

Once a set of labelled hashtags is generated, we use them to classify tweets into opinion categories by counting number of hashtags for each label and assign the most common label to the tweet. The following 8 opinion categories are generated: Pro-Clinton, Anti-Trump, Support-Clinton (with Pro-Clinton and Anti-Trump two most common labels), Pro-Trump, Anti-Clinton, Support-Trump (with Pro-Trump and Anti-Clinton two most common labels), mixed (with one of Pro-Clinton and Anti-Trump and one of Pro-Trump and Anti-Clinton as two most common labels), and unidentified (no labelled hashtag). We use only tweets for the first 6 categories, in total 238,142, for training.

## B. Relative opinion measure at the word and user levels

As the opinion-oriented word embedding measures relative opinion at the word level, we examine the trained embeddings for the labelled hashtags to evaluate model performance at the word level. Fig. 2 plots a two-dimensional Multi-Dimensional Scaling (MDS) representation of the hashtags in Table 1. The four seed hashtags (#maga for pro-Trump, #imwithher for pro-Hillary, #nevertrump for anti-Trump, and #neverhillary for anti-Clinton) highlighted by black circles showing a bipolarized pattern. Fig. 2 shows that the two groups, Pro-Trump/Anti-Clinton and the Pro-Clinton/Anti-Trump, of hashtags are clearly separated from one another (the dash line shows a rough dividing boundary for the two groups), while within each group the two categories of hashtags are mixed together. It indicates that our trained embeddings correctly perform with learned opinion orientations.

Furthermore, the relative opinion measure at the user level can be obtained by aggregating word embeddings first to the tweet document level then to the user level. To evaluate the performance of user-level relative opinion measure, we selected the user accounts for the two candidates, Donald Trump and Hillary Clinton, whose opinion orientations are publicly known. Fig. 3 shows a two-dimensional MDS representation of the aggregated opinion embeddings for the two candidates and the word embeddings of the four seed hashtags as references. The four seed hashtags are bipolarized between Pro-Trump/Anti-Clinton and the Pro-Clinton/Anti-Trump, as they are trained embeddings (in-sample prediction). The two candidates' user accounts can be reasonably separated and they are toward the expected opinion orientations respectively according to their distances to the four opinion extremes identified by the seed hashtags. As the two candidates' opinion embeddings are inferred (out-of-sample prediction), their separation is not as extreme as the bipolarization of seed hashtags. Rather, they show some similarity, because both user accounts have posted tweets that include hashtags belonging to the opposing opinion categories for the purpose of attack or criticizing.

**Fig 2. Two-dimensional MDS representation of the word embeddings for the hashtags in Table 1.** Red: Pro-Trump, Orange: Anti-Clinton, Blue: Pro-Clinton, Purple: Anti-Trump.

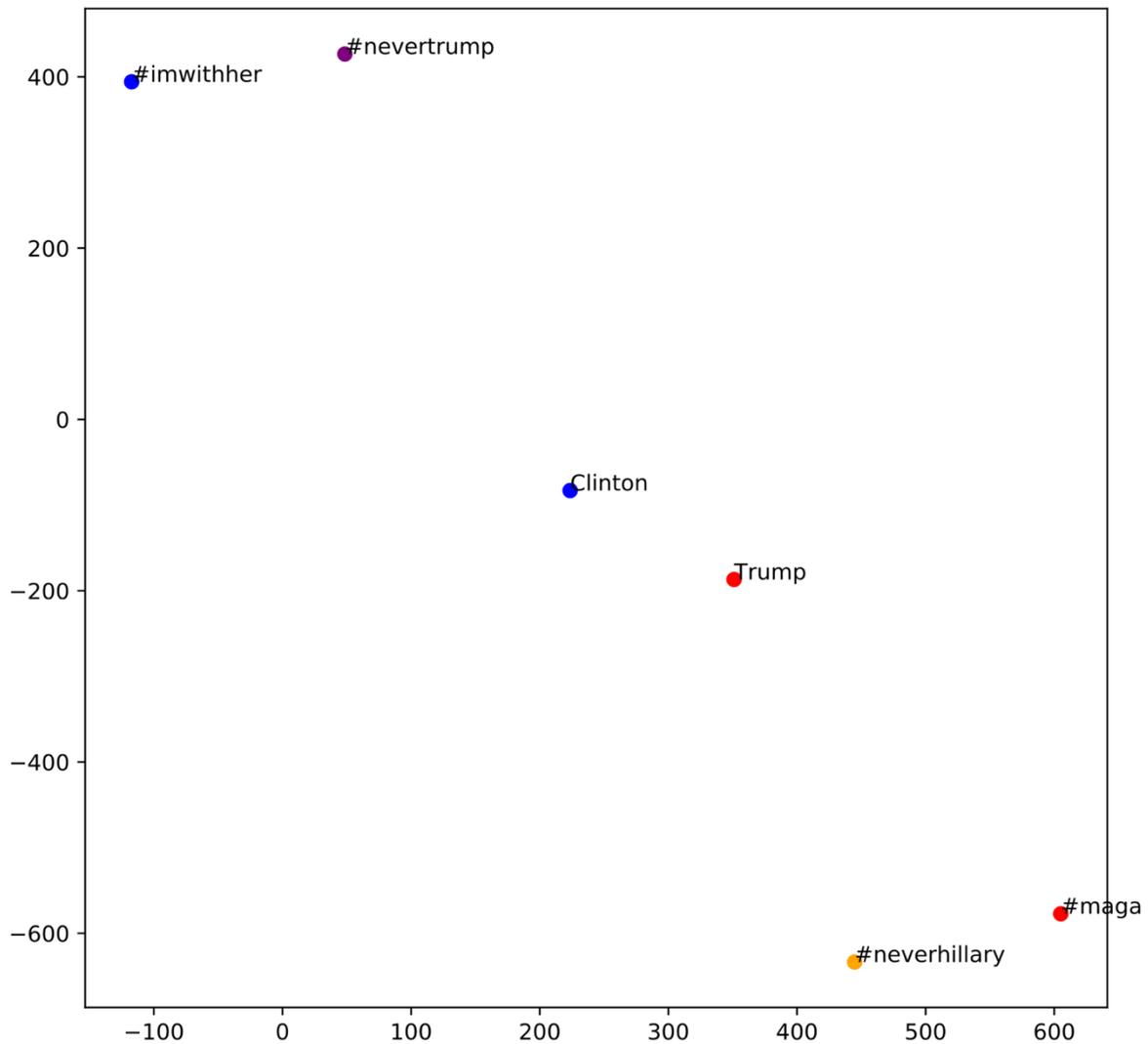

**Fig 3. Clusters of hashtags.** Red: Pro-Trump, Orange: Anti-Clinton, Blue: Pro-Clinton, Purple: Anti-Trump.

# References

1. Martinez-Romo J, Araujo L, Borge-Holthoefer J, Arenas A, Capitán JA, Cuesta JA. Disentangling categorical relationships through a graph of co-occurrences. Phys Rev E. 2011;84: 046108. doi:10.1103/PhysRevE.84.046108

2. Bovet A, Morone F, Makse HA. Validation of Twitter opinion trends with national polling aggregates: Hillary Clinton vs Donald Trump. Sci Rep. 2018;8: 8673. doi:10.1038/s41598-018-26951-y

3. Raghavan UN, Albert R, Kumara S. Near linear time algorithm to detect community structures in large-scale networks. Phys Rev E. 2007;76: 036106. doi:10.1103/PhysRevE.76.036106